\documentclass{SCIS2026}
\usepackage[utf8]{inputenc}
\usepackage{textcomp}

\usepackage{amsmath,amsfonts}

\usepackage{graphicx}
\usepackage[table]{xcolor}
\usepackage{tikz}
\usetikzlibrary{shapes,arrows,positioning}
\usetikzlibrary{arrows.meta, shadows, positioning, shapes.arrows}
\usepackage{pgfplots}
\pgfplotsset{compat=1.18}
\usepackage[edges]{forest}

\usepackage{array}
\usepackage{booktabs}
\usepackage{multirow}
\usepackage{tabularx}
\usepackage{makecell}
\usepackage{stfloats}
\usepackage{nicematrix}

\usepackage{algorithmic}
\usepackage{algorithm}

\usepackage{subfig}

\usepackage{url}
\usepackage{verbatim}
\usepackage{enumitem}

\usepackage{cite}

\usepackage{hyperref}
\usepackage{cleveref}

\makeatletter
\renewcommand\@makefnmark{\hbox{\@textsuperscript{\normalfont\@thefnmark}}}
\makeatother
\crefname{figure}{Figure}{Figures}
\crefname{table}{Table}{Tables}
\crefname{section}{Section}{Sections}

\newif\ifreleaseversion
\releaseversionfalse
\newcommand{\release}{\releaseversiontrue}
\newcommand{\fix}[1]{\ifreleaseversion#1\else{\color{green!50!black}#1}\fi}
\release

\makeatletter
\newcommand{\ie}{\textit{i.e.}\@ifnextchar.{\!\@gobble}{}}
\newcommand{\eg}{\textit{e.g.}\@ifnextchar.{\!\@gobble}{}}
\newcommand{\etc}{etc\@ifnextchar.{}{.\@}}
\makeatother

\begin{document}

\ArticleType{REVIEW}
\Year{2025}
\Month{January}
\Vol{68}
\No{1}
\DOI{}
\ArtNo{}
\ReceiveDate{}
\ReviseDate{}
\AcceptDate{}
\OnlineDate{}
\AuthorMark{}
\AuthorCitation{}

\title{Never compromise with vulnerabilities: a comprehensive survey on AI governance}
      {A comprehensive survey on AI governance}

\makeatletter

\def\@address{}
\def\@addressline#1#2{\textsuperscript{\rm#1}\hspace{0.15em}#2}
\renewcommand*{\address}[2][]{
  \ifx\@address\@empty \gdef\@address{\@addressline{#1}{#2}}
  \else \g@addto@macro\@address{\tabularnewline\@addressline{#1}{#2}} \fi
}
\newcommand*{\sameaddress}[2][]{
  \ifx\@address\@empty \gdef\@address{\@addressline{#1}{#2}}
  \else \g@addto@macro\@address{\quad\quad\@addressline{#1}{#2}} \fi
}

\newcommand\maketitlefirstpart{
    \bgroup%
    \ifthenelse{\pageref{LastPage}=1}
        {\def\@pagerank{\@ArtNo}}{\def\@pagerank{\@ArtNo:1--\@ArtNo:\pageref{LastPage}}}
    \thispagestyle{empty}%
    \begin{picture}(0,0)
        \put(165,23){\journalnamesize\bf \textcolor[rgb]{0.01,0.12,0.67}{SCIENCE CHINA}}
        \put(167,8){\journalnamesize \textcolor[rgb]{0.01,0.12,0.67}{Information Sciences}}
        \put(0,0){\color[rgb]{0.01,0.12,0.67}{\rule{\linewidth}{0.5pt}}}
        \ifx\@DOI\@empty\else
            \put(0,7){\makebox[\textwidth][r]{\href{http://crossmark.crossref.org/dialog/?doi=\@DOI&domain=pdf&date_stamp=\today}{\includegraphics{Print-CrossMark.eps}}}}
            \put(0,-9){\makebox[\textwidth][r]{\vbox{\hbox{\headerfootersize \@Month~\@Year,~Vol.~\@Vol,~Iss.~\@No,~\@pagerank}}}}
            \put(0,-18){\makebox[\textwidth][r]{\vbox{\hbox{\headerfootersize \href{https://doi.org/\@DOI}{https://doi.org/\@DOI}}}}}
            \ifx\@luntan\@empty\else \put(0,-90){\makebox[\textwidth][r]{\includegraphics{kxqylt.eps}}} \fi
            \ifx\@oatag\@empty \put(0,-660){\makebox[\textwidth][l]{\headerfootersize\sf \@copyright}}
            \else \put(0,-660){\makebox[\textwidth][l]{\headerfootersize\sf \@oacopyright}} \fi
            \put(0,-660){\makebox[\textwidth][r]{\headerfootersize\sf \@website}}
        \fi
    \end{picture}
    \begin{picture}(0,0) \rm
    \put(0,-17){\makebox[\textwidth][l]{\bf \textcolor[rgb]{0.01,0.12,0.67}{{\LARGE\bfseries\raisebox{2pt}{.}}~\@ArticleType~{\LARGE\bfseries\raisebox{2pt}{.}}}}}
    \ifx\@ArticleType\editorial\else \put(1,-30){\@SpecialTopic} \fi
    \end{picture}
    \noindent\vskip 18mm
    \begin{center}\titlesize\bfseries\@title\end{center}
    \ifx\@ArticleType\editorial\else
      \begin{center}
        \authorsize \rm
        \begin{tabular}{p{0.9\textwidth}}
          \centering
          \linespread{1.3}\selectfont
          Yuchu JIANG\textsuperscript{1}\footnotemark[2]\footnotemark[3], Jian ZHAO\textsuperscript{1}\footnotemark[2]\footnotemark[1], Yuchen YUAN\textsuperscript{1}\footnotemark[2], Tianle ZHANG\textsuperscript{1}\footnotemark[2], Yao HUANG\textsuperscript{2}\footnotemark[2], Yanghao ZHANG\textsuperscript{3}, Jiaqi LIU\textsuperscript{1,4}, Haijun SONG\textsuperscript{1,5}, Yan WANG\textsuperscript{1,6}, Yanshu LI\textsuperscript{1,7}, Xizhong GUO\textsuperscript{1,7}, Huilin ZHOU\textsuperscript{1,8}, Yusheng ZHAO\textsuperscript{1,8}, Jun ZHANG\textsuperscript{1,7}, Zhi ZHANG\textsuperscript{9}, Xiaojian LIN\textsuperscript{10}, Yixiu ZOU\textsuperscript{11}, Haoxuan MA\textsuperscript{12}, Yuhu SHANG\textsuperscript{5}, Yuzhi HU\textsuperscript{13}, Keshu CAI\textsuperscript{13}, Ruochen ZHANG\textsuperscript{2}, Boyuan CHEN\textsuperscript{14}, Yilan GAO\textsuperscript{15}, Ziheng JIAO\textsuperscript{15}, Yi QIN\textsuperscript{16}, Shuangjun DU\textsuperscript{7}, Tong XIAO\textsuperscript{16}, Zhekun LIU\textsuperscript{16}, Yu CHEN\textsuperscript{16}, Xuankun RONG\textsuperscript{17}, Rui WANG\textsuperscript{16}, Yejie ZHENG\textsuperscript{18}, Zhaoxin FAN\textsuperscript{2}, Murat SENSOY\textsuperscript{19}, Hongyuan ZHANG\textsuperscript{20}, Pan ZHOU\textsuperscript{21}, Lei JIN\textsuperscript{5}, Hao ZHAO\textsuperscript{10}, Xu YANG\textsuperscript{12}, Jiaojiao ZHAO\textsuperscript{9}, Jianshu LI\textsuperscript{22}, Joey Tianyi ZHOU\textsuperscript{23}, Zhi-Qi CHENG\textsuperscript{13}, Longtao HUANG\textsuperscript{24}, Zhiyi LIU\textsuperscript{20}, Zheng ZHU\textsuperscript{25}, Jianan LI\textsuperscript{11}, Gang WANG\textsuperscript{26}, Qi LI\textsuperscript{7}, Xu-Yao ZHANG\textsuperscript{7}, Yaodong YANG\textsuperscript{14}, Mang YE\textsuperscript{17}, Wenqi REN\textsuperscript{27}, Zhaofeng HE\textsuperscript{5}, Hang SU\textsuperscript{10}, Rongrong NI\textsuperscript{16}, Liping JING\textsuperscript{16}, Xingxing WEI\textsuperscript{2}, Junliang XING\textsuperscript{10}, Massimo ALIOTO\textsuperscript{28}, Shengmei SHEN\textsuperscript{29}, Petia RADEVA\textsuperscript{30}, Dacheng TAO\textsuperscript{31}, Ya-Qin ZHANG\textsuperscript{10}, Shuicheng YAN\textsuperscript{28} \& Xuelong LI\textsuperscript{1}\footnotemark[1]
        \end{tabular}
      \end{center}
      \footnotetext[1]{Corresponding authors: Xuelong Li and Jian Zhao (\{xuelong\_li, zhaoj90\}@chinatelecom.cn).}
      \footnotetext[2]{Equal contribution: Yuchu Jiang (kamichanw@seu.edu.cn), Jian Zhao, Yuchen Yuan, Tianle Zhang (\{zhaoj90, yuanyc2, zhangtl15\}@chinatelecom.cn) and Yao Huang (y\_huang@buaa.edu.cn).}
      \footnotetext[3]{Work done during an internship at TeleAI.}
      \vspace*{-2.5ex}
      \begin{center}\headerfootersize{\it \begin{tabular}[t]{@{}>{\centering\arraybackslash}p{0.98\textwidth}@{}} \@address \end{tabular}} \end{center}
    \fi
    \egroup%
}

\newcommand\maketitleabstractpart{
    \bgroup%
    \thispagestyle{empty}%
    \ifx\@ArticleType\erratum\else\ifx\@ArticleType\editorial\else\ifx\@ArticleType\supplementary\else
      \begin{center}
          \footnotesize
          \ifx\@DOI\@empty \else \ifx\@ReviseDate\@empty
              \ifx\@OnlineDate\@empty{Received \@ReceiveDate/Accepted \@AcceptDate}\else{Received \@ReceiveDate/Accepted \@AcceptDate/Published online \@OnlineDate}\fi
          \else
              \ifx\@OnlineDate\@empty{Received \@ReceiveDate/Revised \@ReviseDate/Accepted \@AcceptDate}\else{Received \@ReceiveDate/Revised \@ReviseDate/Accepted \@AcceptDate/Published online \@OnlineDate}\fi
          \fi\fi
      \end{center}
      \vspace*{-3ex}%
      \begin{center}
          \begin{tabular}{p{0.95\textwidth}}
          \arrayrulecolor{sciscolor}\toprule
              \ifx\@ArticleType\moop\else \ifx\@ArticleType\perspective\else \ifx\@ArticleType\news\else \ifx\@ArticleType\highlight\else \ifx\@ArticleType\letter\else
                \abstractsize\noindent\textcolor{black}{\textbf{Abstract}\quad\ignorespaces\@abstract}\\
                \abstractsize\noindent\textcolor{black}{\textbf{Keywords}\quad\ignorespaces\@keywords}\\ \hline
              \fi\fi\fi\fi\fi
              \abstractsize\noindent\textcolor{black}{\textbf{Citation}\quad\ignorespaces\citationline}\\ \bottomrule
          \end{tabular}
      \end{center}
    \fi\fi\fi
    \egroup%
}

\makeatother
\abstract{The rapid advancement of artificial intelligence (AI) has significantly expanded its capabilities across diverse domains. However, this also introduces complex technical vulnerabilities, such as algorithmic biases and adversarial sensitivity, that can carry significant societal risks, including misinformation, inequity, cybersecurity concerns, physical-world accidents, and declines in public credibility. These challenges underscore the pressing need for AI governance to inform the development and deployment of AI technologies. To meet this need, we propose a comprehensive AI governance framework that integrates both technical and societal dimensions simultaneously. Specifically, we categorize governance into three interconnected aspects: \fix{\textbf{intrinsic security}} (internal system reliability), \fix{\textbf{derivative security}} (external real-world harms), and \fix{\textbf{ethical security}} (value alignment and accountability). Uniquely, we integrate technical methodologies, emerging evaluation benchmarks, and policy perspectives to construct a governance framework that actively supports transparency, accountability, and public trust. Through a systematic review of over 300 references, we identify three critical systematic challenges: (1) the generalization gap, where existing defenses fail to adapt to the evolving threats; (2) evaluation protocols that insufficiently reflect real-world deployment risks; and (3) fragmented regulatory landscapes that produce inconsistent oversight and enforcement. We attribute these failures to a fundamental misalignment in current practices, where governance is treated as an afterthought rather than a foundational design principle. As a result, existing efforts tend to be reactive and piecemeal, falling short in addressing the inherently interconnected nature of technical reliability and societal trust. In response, our study provides a comprehensive landscape analysis and articulates an integrated research agenda that bridges technical rigor with social responsibility. This framework equips researchers, engineers, and policymakers with actionable insights for designing AI systems that not only exhibit performance robustness but also align with ethical imperatives and public trust. \fix{The accompanying repository is available at \url{https://github.com/Tele-EVOL/AI-Governance}.}}


\address[1]{Institute of Artificial Intelligence (TeleAI), China Telecom, Beijing 100032, China}
\address[2]{Beihang University, Beijing 100191, China}
\sameaddress[3]{Imperial College London, UK}
\address[4]{Zhejiang University, Hangzhou 310027, China}

\address[5]{Beijing University of Posts and Telecommunications, Beijing 100088, China}
\sameaddress[6]{University of Edinburgh, UK}
\address[7]{University of Chinese Academy of Sciences, Beijing 100049, China}
\address[8]{University of Science and Technology of China, Hefei 230026, China}

\address[9]{University of Amsterdam, The Netherlands}
\sameaddress[10]{Tsinghua University, Beijing 100084, China}
\address[11]{Beijing Institute of Technology, Beijing 100081, China}
\sameaddress[12]{Southeast University, Nanjing 211189, China}

\address[13]{University of Washington, USA}
\sameaddress[14]{Peking University, Beijing 100871, China}
\address[15]{Northwestern Polytechnical University, Xi'an 710072, China}
\sameaddress[16]{Beijing Jiaotong University, Beijing 100044, China}

\address[17]{Wuhan University, Wuhan 430072, China}
\address[18]{Shanghai Collaborative Innovation Center for AI Social Governance, Shanghai 200240, China}
\sameaddress[19]{Amazon, USA}
\address[20]{International Research Center for Complexity Sciences, Hangzhou International Innovation Institute, Beihang University, Hangzhou 311115, China}

\address[21]{Singapore Management University, Singapore}
\sameaddress[22]{Ant Group, Singapore}
\address[23]{CFAR, Technology and Research (A*STAR), Singapore}
\sameaddress[24]{Alibaba Group, Hangzhou 311121, China}

\address[25]{GigaAI, Beijing 100084, China}
\sameaddress[26]{Academy of Military Medical Sciences, Beijing 100850, China}
\address[27]{Shenzhen Campus of Sun Yat-sen University, Shenzhen 518107, China}

\address[28]{National University of Singapore, Singapore}
\sameaddress[29]{Pensees Singapore, Singapore}
\address[30]{University of Barcelona, Spain}
\sameaddress[31]{Nanyang Technological University, Singapore}

\maketitlefirstpart
\newpage
\maketitleabstractpart

\section{Introduction}

The rapid advancement of artificial intelligence (AI), particularly the emergence of large language models (LLMs), has brought transformative changes across science \cite{eger2025transforming}, industry \cite{chkirbene2024large}, and society \cite{farrell2025large}. Specifically, these models excel in reasoning \cite{huang2023towards}, content generation \cite{chang2024survey}, and decision support \cite{ma2025towards}, enabling a wide range of applications across education, healthcare, law, and public services. This widespread adoption also reflects the growing confidence in AI's potential to augment human capabilities and drive societal progress.

However, the deployment of AI systems at scale has simultaneously surfaced a host of new risks. Unlike conventional vulnerabilities that primarily affect software system functionality, AI-specific risks may manifest in more concerning ways that can undermine public trust and cause widespread harm. For instance, LLMs can be manipulated through prompt injection attacks to bypass safety mechanisms and generate harmful or illegal content \cite{greshake2023not}. Generative models enable the creation of convincing deepfakes for large-scale misinformation campaigns and non-consensual intimate imagery \cite{wu2025survey}, fundamentally challenging our ability to distinguish authentic content from fabricated material. Perhaps most critically, AI hallucinations \cite{huang2025survey} in high-stakes domains can lead to catastrophic outcomes—from erroneous medical diagnoses that endanger patient lives \cite{mei2020artificial} to flawed financial advice that destroys livelihoods. These cases underscore a critical truth that AI risks are no longer theoretical concerns but are already affecting individuals, communities, and institutions in the real world.

In response to these risks, the concept of AI governance has emerged, which encompasses the rules, practices, and technologies that guide the development and deployment of AI throughout its entire lifecycle to be ethically aligned, legally compliant, and socially beneficial \cite{dafoe2018ai,wang2025comprehensive,wei2025responsible}. More importantly, rather than treating safety as a post hoc add-on, AI governance calls for a proactive, integrated approach to managing AI risks \cite{jobin2019global}.

\begin{figure*}[!t]
   \centering
   \includegraphics[width=0.85\textwidth]{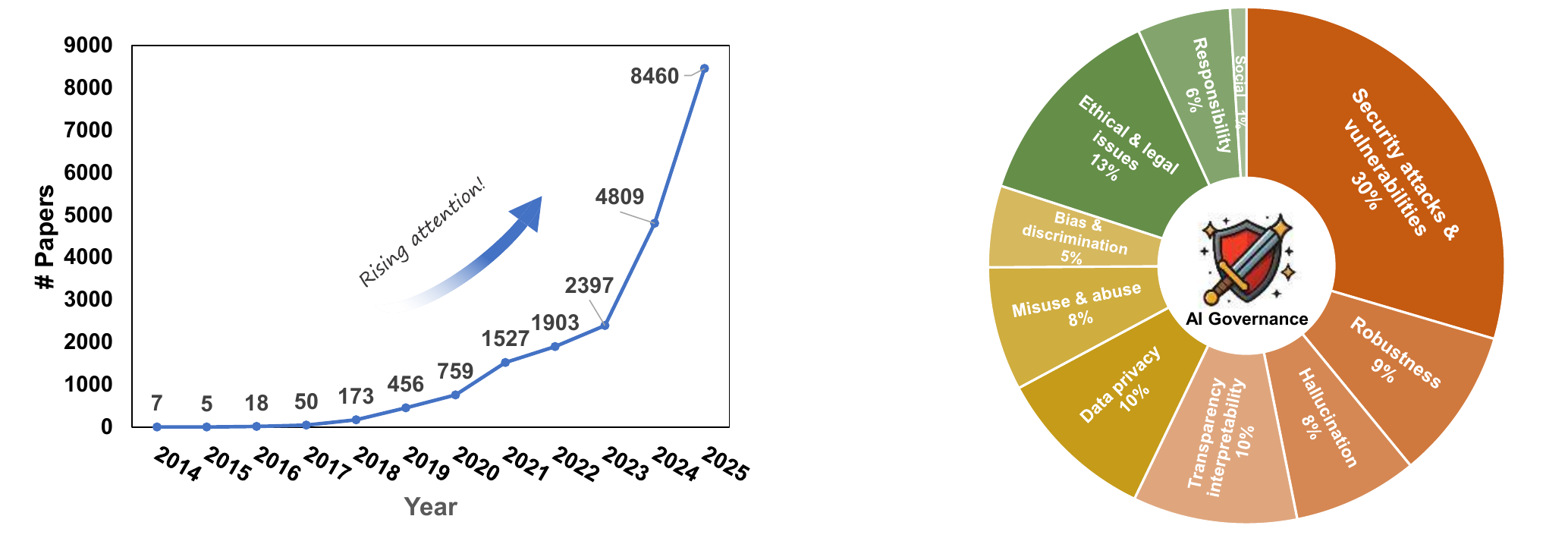}
    \caption{\textbf{Left}: The number of AI governance papers published over the past years. \textbf{Right}: The distribution of research across different dimensions.}
   \label{fig:pub}
\end{figure*}

\begin{figure*}[!ht]
   \centering
   \includegraphics[width=1\textwidth]{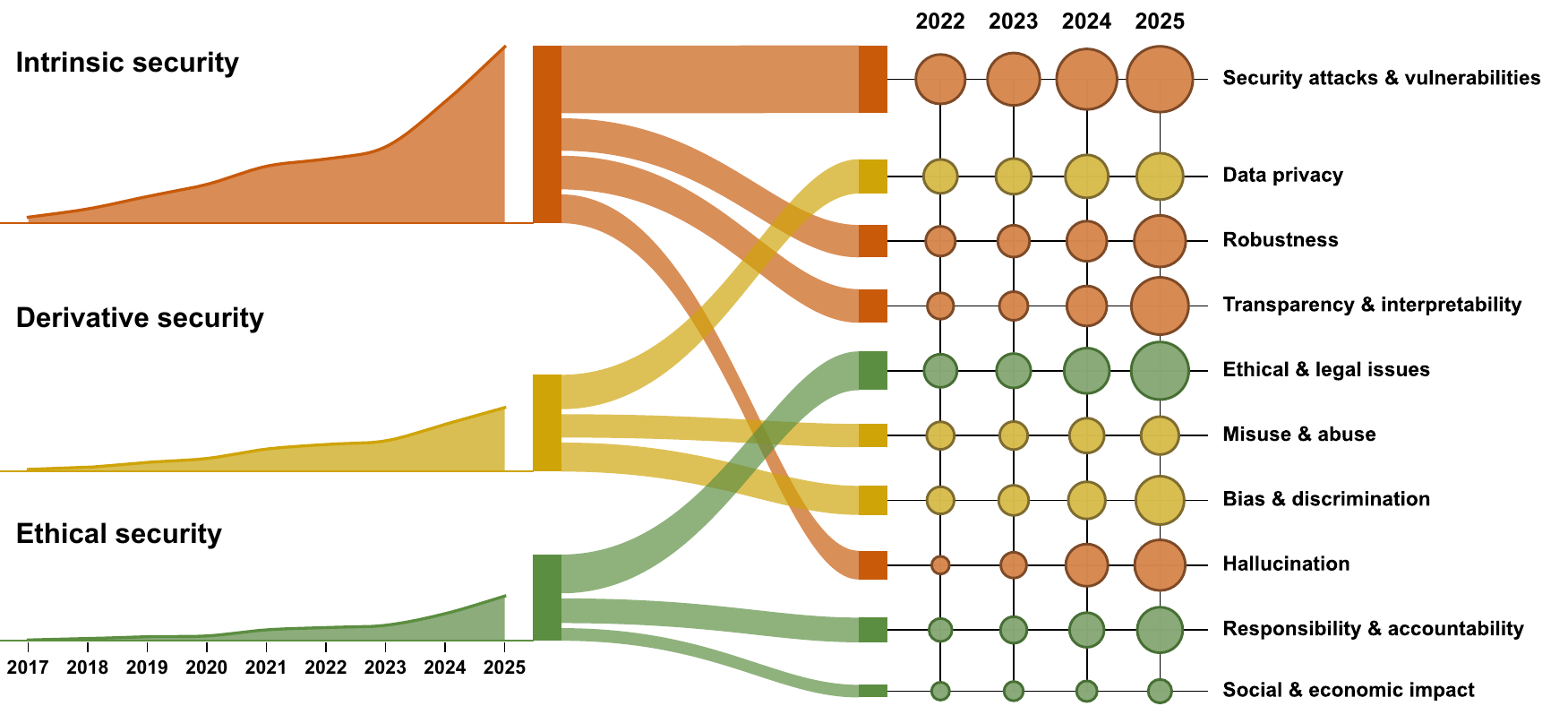}
    \caption{\textbf{Left}: The quarterly trend in the number of published governance papers across different dimensions. \textbf{Middle}: Proportional distribution of intrinsic security, derivative security, and ethical security. \textbf{Right}: The annual trend in the number of governance research papers published on various dimensions, presented in descending order from highest to lowest.}
   \label{fig:pub-stats}
\end{figure*}

Figures~\ref{fig:pub} and~\ref{fig:pub-stats} summarize the publication volume, dimensional distribution, and temporal trends used to characterize this research landscape.

To fully understand the current state and development of AI governance, we have examined the existing research landscape. The temporal distribution of selected publications, as shown in ~\cref{fig:pub}, reveals a clear trend in the evolving landscape of AI governance research. Between 2017 and 2024, there has been a remarkable increase in academic interest. By the end of 2025, the number of related academic papers exceeded 8000, suggesting that the rapid deployment of AI in real-world applications has prompted urgent discussions about its governance.

\begin{figure*}[!t]
   \centering
   \includegraphics[width=\textwidth]{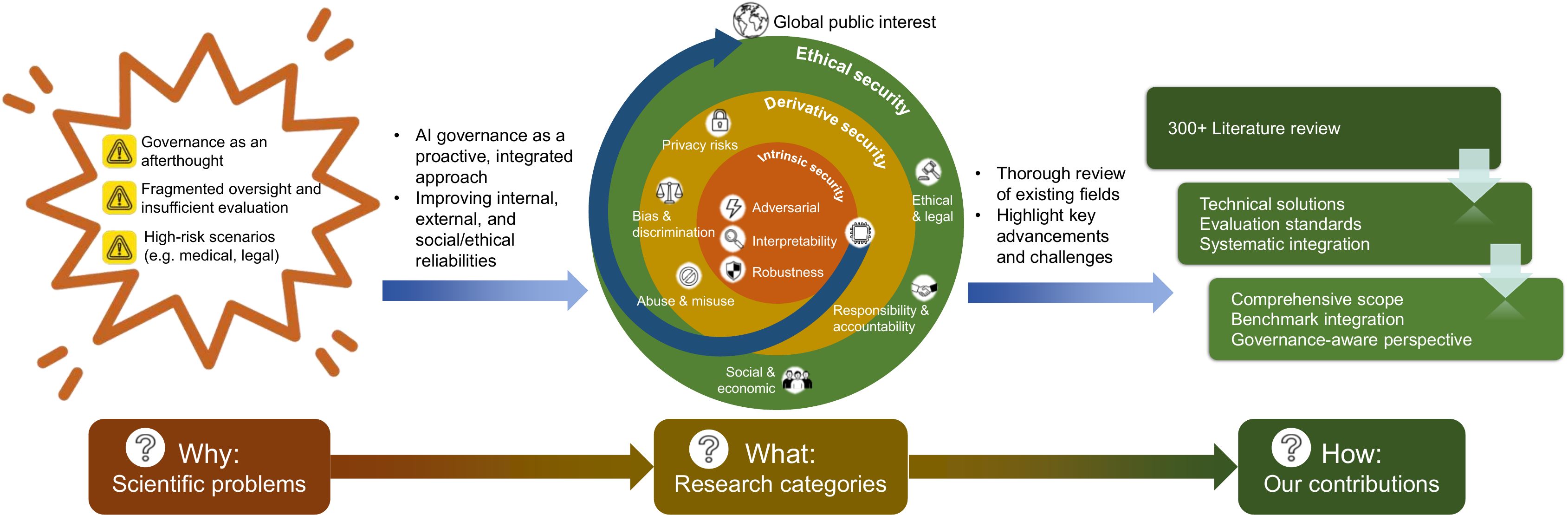}
    \caption{A conceptual overview of the AI governance framework, illustrating the key motivations, research categories, and contributions of this survey.}
   \label{fig:structure}
\end{figure*}

Figure~\ref{fig:structure} presents the three-layer governance framework and its relationship to the motivations and contributions of this survey.

Despite this rapid growth in research volume, the field still lacks a systematic and technically grounded synthesis that bridges different domains. Existing studies \cite{deng2025ai} tend to isolate technical safety from broader governance considerations or narrowly focus on specific risks such as fairness or adversarial robustness without offering a unifying framework. A parallel body of scholarship, rooted mainly in ethics and legal studies, offers high-level normative analyses but rarely engages with the emerging toolkit of empirical evaluation methods, standardized benchmarks, and system-level defenses \cite{kaur2022trustworthy}.
In recent years, several surveys have made important progress in trustworthy AI, AI security, privacy, and generative model safety. However, as summarized in \cref{tab:survey_comparison}, existing surveys differ significantly in scope and focus: some concentrate on a single technical area such as privacy or security, while others address governance and regulation to a certain extent. Nevertheless, they generally lack systematic integration of benchmark analysis, multimodal coverage, and a unified taxonomy.
Consequently, an integrative survey is urgently needed to map the full landscape of AI-governance challenges and to situate them within the rapidly evolving architecture of contemporary AI systems.

\begin{table*}[t]
\centering
\caption{\textbf{Comparison with representative surveys.}}
\label{tab:survey_comparison}

\resizebox{\textwidth}{!}{
\small
\begin{tabular}{p{3cm}ccccccc}
\toprule
\textbf{Survey}
& \textbf{Privacy}
& \textbf{Governance \&}
& \textbf{Benchmark}
& \textbf{Multimodal}
& \textbf{Unified} \\

& \textbf{}
& \textbf{regulation}
& \textbf{analysis}
& \textbf{coverage}
& \textbf{taxonomy} \\
\midrule

Kaur et al.~\cite{kaur2022trustworthy}
& \ding{51}
& \ding{51}
& \ding{53}
& \ding{53}
& \ding{53} \\

Rahman et al.~\cite{rahman2023security}
& \ding{51}
& \ding{53}
& \ding{53}
& \ding{53}
& \ding{53} \\

Chen and Babar~\cite{chen2024security}
& \ding{51}
& \ding{53}
& \ding{53}
& \ding{53}
& \ding{53}  \\

Golda et al.~\cite{golda2024privacy}
& \ding{51}
& \ding{51}
& \ding{53}
& \ding{51}
& \ding{53} \\

Huang et al.~\cite{huang2025survey}
& \ding{53}
& \ding{53}
& \ding{51}
& \ding{51}
& \ding{53} \\

Our survey
& \ding{51}
& \ding{51}
& \ding{51}
& \ding{51}
& \ding{51}  \\

\bottomrule
\end{tabular}
}
\end{table*}

To address this demand, we propose this comprehensive survey, which aims to provide a systematic examination of AI governance that could serve as a reference for researchers, developers, and policymakers working to ensure AI systems are robust, accountable, and aligned with public interest. Specifically, our work seeks to answer three key questions: (1) \textbf{Why is it urgent to investigate AI governance?} We identify the research gap where governance is primarily treated as an afterthought rather than a core design principle, leading to fragmented oversight and insufficient evaluation in existing defenses. This motivates our survey, which situates AI governance as an essential foundation for trustworthy AI. (2) \textbf{How can technical methods, benchmarks, and policy perspectives be integrated into a unified governance framework?} We define a unified governance framework encompassing three key dimensions: intrinsic security (\eg, adversarial robustness, hallucination, interpretability), derivative security (\eg, privacy, bias, misuse), and ethical security (\eg, legal norms, accountability mechanisms, emerging ethical concerns). Through this taxonomy, we review technical and societal risks in a coherent and structured manner. (3) \textbf{What open challenges and governance implications can be synthesized from the existing literature?} We conduct a systematic review of over 300 references, analyze the representative benchmarks and evaluation metrics across vision, language, and multimodal systems, compare the strengths and weaknesses of existing methodologies, and synthesize open challenges and future research directions. This multidimensional review offers actionable insights for researchers, engineers, and policymakers seeking to develop AI systems that are not only robust and reliable but also socially responsible and ethically aligned.

For a better presentation of our survey, we structure it according to the following principles to ensure greater clarity and logical organization:
\begin{itemize}
    \item \textbf{Taxonomy.}
As illustrated in ~\cref{fig:structure}, we organize AI governance into three core pillars: intrinsic security, derivative security, and ethical security. The taxonomy follows an outward-moving governance lens, moving from risks that can be assessed mainly at the level of the AI system, to risks that emerge through deployment and interaction, and finally to risks that require institutional and legal response. \emph{Intrinsic security} concerns model-centered reliability, including whether a system remains stable under perturbations or distribution shifts, produces faithful and grounded outputs, and exposes sufficient information for inspection. It therefore covers adversarial vulnerability, robustness, hallucination, and interpretability. \emph{Derivative security} concerns externally manifested harms that arise when models, data, outputs, users, and deployment contexts interact. Privacy, bias, and misuse belong to this layer because their governance significance is realized through privacy leakage, discriminatory treatment, or harmful use affecting external parties, even when their causal mechanisms may originate in training data or model parameters. \emph{Ethical security} concerns rights, responsibilities, legal obligations, accountability, and socioeconomic impacts that cannot be resolved solely by modifying a single model or deployment. We use ``security'' in this third layer not to equate ethics with conventional cybersecurity, but to emphasize that failures in rights protection, responsibility allocation, legal compliance, and social trust can undermine the security of social and institutional systems.

    \item \textbf{Organization.} We provide a systematic analysis of each pillar through a consistent analytical framework. For each sub-dimension, we follow a structured approach that typically includes problem definition, followed by relevant aspects such as risk analysis, attack methodologies (where applicable), evaluations (when available), and mitigation strategies. This flexible yet uniform structure accommodates the diverse nature of governance challenges while maintaining systematic coverage across different domains.
\end{itemize}
\hypertarget{Intrinsic security}{}

\vspace{-0.3em}

\section{Intrinsic security} \label{sec:intrinstic}

Intrinsic security refers to the model-centered properties that determine whether an AI system behaves reliably, stably, and transparently before external safeguards or institutional governance mechanisms are considered. It focuses on failures whose primary object of analysis is the model itself, including its behavior, internal mechanisms, factual reliability, and inspectability, while recognizing that such failures may lead to severe downstream consequences in real-world applications. As AI systems become deeply integrated into high-stakes domains, weaknesses in adversarial resistance, out-of-distribution generalization, factuality and faithfulness, and model transparency can undermine both system performance and public trust. This section systematically examines four core dimensions of intrinsic security—adversarial vulnerability, robustness, hallucination, and interpretability—each exposing a distinct facet of model fragility.

\hypertarget{Adversarial vulnerability}{}
\subsection{Adversarial vulnerability} \label{sec:attcks}

\subsubsection{Problem definition}

Adversarial vulnerabilities refer to the susceptibility of AI models to carefully crafted inputs that induce incorrect or harmful behaviors, compromising system integrity, confidentiality, and availability\cite{szegedy2013intriguing}. These attacks span white-box\cite{carlini2017towards} and black-box settings\cite{papernot2017practical}, and have evolved from simple $\ell_p$-norm perturbations\cite{goodfellow2014explaining} to semantic, physical-world, and multimodal manipulations\cite{bhattad2019unrestricted}. Emerging threats, such as jailbreak prompts in LLMs\cite{yi2024jailbreak}, further bypass safety mechanisms. Despite progress in adversarial training\cite{jiang2025safety}, detection\cite{li2025road}, purification\cite{nie2022diffusion}, and alignment-based defenses\cite{ouyang2022training}, many defenses remain brittle under adaptive attacks\cite{athalye2018obfuscated}, highlighting the need for generalizable mitigation and standardized evaluation frameworks.

\vspace{-0.8em}
\subsubsection{Adversarial attacks}
Adversarial attacks exploit model vulnerabilities by crafting inputs that induce incorrect or harmful outputs. The existing mainstream methods can be categorized into three types:
\begin{itemize}
    \item \fix{\textbf{White-box attacks.}} The adversary has full knowledge of the target model (architecture, parameters, training data, \etc.) \cite{dong2018boosting}. The attacker can compute gradients and craft perturbations directly. This setting yields powerful attacks but is less realistic in deployed systems.

    \item \fix{\textbf{Black-box attacks.}} Adversaries cannot access model internals and must either query the model for outputs or exploit transferability from other models. These attacks, which better reflect real-world conditions, are generally categorized as transfer-based\cite{papernot2017practical,fang2024strong} or query-based\cite{andriushchenko2020square}.

    \item \fix{\textbf{Emerging LLM attacks.}} Jailbreak prompts~\cite{zou2023universal, yi2024jailbreak} bypass safety filters using crafted textual inputs, while backdoor attacks \cite{liu2017trojan} implant triggers during training to induce conditional misbehavior.
\end{itemize}

\subsubsection{Adversarial defenses}
Defenses have evolved from input preprocessing to integrated robustness strategies.
\begin{itemize}
    \item \fix{\textbf{Vision-model defenses.}} Adversarial training remains the most effective, with efficient variants like AGAT\cite{wu2022towards}  and ARD-PRM\cite{mo2022adversarial} designed for ViTs. Detection-based methods identify anomalous inputs via feature artifacts\cite{li2025road}, while purification techniques\cite{khalili2024lightpure} attempt to remove perturbations using diffusion models or lightweight filters.

    \item \fix{\textbf{LLM defenses.}} Alignment via RLHF provides foundational safety, but must be reinforced with runtime defenses such as input perplexity filters\cite{xiong2024defensive}, circuit breakers\cite{zou2024improving}, or ensemble-based rewriting frameworks like AutoDefense\cite{ye2024we}, MoGU\cite{zhang2024adversarial}. These defenses mitigate jailbreaks and toxic outputs in open-ended generation.

    \item \fix{\textbf{VLM defenses.}} Prompt-based defenses adapt inputs or prompts with limited parameter updates. Representative methods include APT\cite{li2024one}, Defense-Prefix\cite{azuma2023defense}, FAP\cite{zhou2024few}, and TAPT\cite{wang2024tapt}, which improve robustness by optimizing textual prompts or jointly adapting visual and textual signals.

    \item \fix{\textbf{Cross-modal detection and MLLM defenses.}} Detection in VLMs can leverage cross-modal consistency. Tools such as MirrorCheck\cite{fares2024mirrorcheck} and AdvQDet\cite{wang2024advqdet} identify mismatches between visual and textual signals or suspicious query patterns. For generative MLLMs, jailbreak defenses such as JailGuard \cite{zhang2023mutation}, GuardMM\cite{sharma2024defending}, and MLLM-Protector\cite{pi2024mllm} combine input inspection, reasoning traceability, and output filtering.
\end{itemize}

\subsubsection{Benchmarks}

To support standardized evaluation and facilitate future research,~\cref{tab:robustness_type} presents representative benchmarks commonly used to assess adversarial robustness. Centered on adversarial robustness, these benchmarks test models under adversarially sourced or framed inputs rather than \textsc{iid} noise. ANLI \cite{nie2020adversarial_nli} uses a human--model iterative loop to craft hard NLI counterexamples that reveal brittle lexical heuristics, measuring robust accuracy under human-generated attacks. AdvGLUE \cite{wang2021adversarial_glue} scales this to a multi-task GLUE setting with semantically preserved adversarial edits and human filtering, enabling per-task robustness drops (clean$\rightarrow$adv) and exposing defense overfitting to attack families. TruthfulQA \cite{lin2022truthfulqa} probes a safety-critical facet of adversarial robustness via belief-aligned prompts designed to trigger human misconceptions; robustness is captured by truthful accuracy and calibration, complementing token-space attacks with content-level adversaries. In multimodal settings, JailbreakV\cite{luo2024jailbreakv} treats jailbreaks as adversarial attacks on MLLMs using text-only and image-conditioned prompts; the key metric is attack success rate (ASR), quantifying how reliably prompts bypass safety while trading off utility/refusal. Although framed around distribution shift, OODRobustBench \cite{li2024oodrobustbench} is directly relevant: it tests whether adversarially trained or ``robust'' models retain robustness when both data and threat models shift, urging reports of robust accuracy under shift rather than only in-distribution outcomes. Finally, broad capability suites---SEED-Bench \cite{Li_2024_CVPR} for vision--language and AIR-Bench \cite{yang2024airbench} for audio--language---offer objective MCQ backbones and diverse inputs on which to layer adversarial variants, enabling cross-modal transfer analysis and separating competence from adversarial degradation. Taken together, \cite{nie2020adversarial_nli,wang2021adversarial_glue,lin2022truthfulqa,luo2024jailbreakv,li2024oodrobustbench,Li_2024_CVPR,yang2024airbench} support a practical recipe: combine human-crafted natural adversaries (ANLI) with systematic multi-task perturbations (AdvGLUE), add security jailbreak stress for MLLMs (JailbreakV; report ASR and utility/refusal), verify transfer under shift (OODRobustBench), and use SEED/AIR as scaffolds; always report clean versus adversarial performance, ASR, and calibration to expose utility--robustness trade-offs.

\subsubsection{Synthesis and governance implications}

Adversarial vulnerability research has moved from attack-specific defenses toward layered protection, including training-time robustness, runtime filtering, input purification, and cross-modal consistency checks. Yet the surveyed methods reveal a persistent generalization gap: defenses optimized for fixed attack settings often degrade against adaptive, unseen, or cross-modal threats. This creates a practical trade-off between computationally intensive robustness mechanisms and lighter, more deployable defenses with narrower coverage.

Adversarial defense should therefore be translated from a purely technical robustness target into a lifecycle governance requirement. Before deployment, high-risk AI systems should undergo threat modeling, adaptive red-teaming, and benchmark-based stress testing under both white-box and black-box assumptions. During deployment, organizations should maintain prompt-injection logs, attack-success-rate monitoring, incident escalation procedures, and patch-management workflows. At the policy level, adversarial robustness evidence can support procurement rules, third-party certification, and sector-specific safety standards by defining minimum acceptable robustness thresholds and documentation requirements for model updates.
\hypertarget{Robustness}{}
\vspace{-0.5em}
\subsection{Robustness} \label{sec:robustness}

\subsubsection{Problem definition}

In the context of intrinsic security, robustness refers to an AI model’s ability to maintain reliable performance under input variations outside the training distribution, encompassing both adversarial robustness against worst-case, human-crafted perturbations and natural robustness (or non-adversarial robustness) to benign but unpredictable distribution shifts\cite{gojic2023non}. Unlike adversarial attacks, such shifts—caused by factors like lighting changes, occlusions, or dialect differences—arise naturally in real-world data and can severely degrade model accuracy\cite{wu2023toward}.

\subsubsection{Existing methods}

Improving natural in-distribution and out-of-distribution (OOD) generalization has motivated a range of training methodologies across modalities. Below, we outline key strategies and their trade-offs.

\begin{itemize}

    \item \fix{\textbf{Data augmentation and corruption simulation.}} A straightforward approach to enhance robustness is to expose models to input variations during training. Techniques such as AugMix~\cite{hendrycks2019augmix} and DeepAugment synthesize diverse corruptions (\eg, noise, blur, color shifts, weather effects), compelling models to learn more invariant and generalizable features. These augmentations significantly improve performance on benchmarks like ImageNet-C~\cite{hendrycks2019benchmarkingCorruptions}. Artificial augmentation has been shown to rival the benefits of scaling up datasets~\cite{modas2022fewprimitives}. However, its effectiveness is often bounded to the perturbation types it simulates and may not generalize to unseen domains or contextual shifts~\cite{kireev2021adversarial}.

    \item \fix{\textbf{Domain generalization and adaptation.}} Domain generalization aims to train models that perform well on unseen domains without access to target-domain data during training. Methods include learning domain-invariant representations, distributionally robust optimization, and meta-learning. For example, in biomedical signals, domain adaptation (\eg, adversarial alignment or pretraining on diverse hospitals) has enabled ECG/EEG models to generalize across sensors and institutions~\cite{ballas2023biodg}. While such strategies improve transfer to related domains, they often falter under severe domain shifts.

    \item \fix{\textbf{Self-supervised and contrastive learning.}} Self-supervised pretraining, particularly contrastive methods (\eg, SimCLR~\cite{chen2020simclr}, MoCo), has demonstrated strong robustness by encouraging invariance to data augmentations and capturing higher-level semantics. In NLP, similar pretraining objectives (\eg, masked language modeling in BERT) confer resilience to syntactic variation. In biosignals, frameworks like BENDR~\cite{kostas2021bendr} applied contrastive learning to EEG data and improved generalization across datasets. Speech SIMCLR~\cite{jiang2020speechsimclr} and wav2vec2.0~\cite{xiao2021wav2vec2} illustrate that contrastive/self-supervised objectives benefit time-series/speech data. Large-scale models like CLIP~\cite{radford2021clip} benefit from diverse, uncurated web data, leading to notable zero-shot robustness. Broad reviews and surveys~\cite{jaiswal2020survey} affirm that self-supervision on varied corpora acts as implicit augmentation. More recent ECG/EEG-specialized work—such as MAEEG~\cite{chien2022maeeg} and scaling ECG representation learning~\cite{avramidis2023ecgssl}—extend these insights to biomedical signals.
    Self-supervision on sufficiently diverse corpora may partially act as an implicit augmentation by encouraging invariance to certain perturbations and semantic variations, although the resulting robustness gains depend heavily on the data distribution, pretraining objective, and downstream task.

\end{itemize}

Other strategies for improving robustness include test-time adaptation\cite{liang2023survey_tta} (which adapt the model at inference using unlabeled test inputs), robust architecture design\cite{wang2025robin} (\eg, convolutional layers invariant to small transformations or vision transformers stable to perturbations), ensemble methods\cite{wang2024garrison} (which enhance robustness and uncertainty estimation by aggregating predictions), and adversarial training, though its impact on natural robustness is mixed. These methods have complementary strengths, and combinations—such as data augmentation with self-supervised pre-training or domain generalization with episodic training followed by test-time adaptation—often yield the best results. To evaluate progress, benchmarks beyond in-distribution accuracy (\eg, ImageNet-A for “natural adversarial” images, ImageNet-O for OOD detection, and WILDS 2.0 with unlabeled adaptation data) have become essential for diagnosing and quantifying robustness under distribution shifts.

\subsubsection{Benchmarks}

To support standardized evaluation and facilitate future research,~\cref{tab:robustness_type} presents representative benchmarks commonly used to assess model robustness (\ie, natural adversarial, natural corruption, natural variation, and OOD). For natural corruptions and perturbations, ImageNet-C\cite{hendrycks2019natural_adversarial} standardize evaluation via mean Corruption Error (mCE) for 15 corruption types and stability metrics for perturbation sequences—mean Flip Rate (mFR) and mean Top-5 Distance (mT5D)—revealing that much of the apparent robustness gains track clean-set accuracy while prediction stability under small input changes remains fragile. Pushing beyond synthetic distortions, Natural Adversarial Examples \cite{hendrycks2019benchmarkingCorruptions} curate ImageNet-A (naturally occurring hard images), on which standard classifiers collapse, underscoring brittleness to real-world edge cases. Natural variation in language-vision grounding is captured by VQA-Rephrasings \cite{shah2019arxiv_vqa_rephrasings}, which adds three human rephrasings per question and introduces a consensus-based robustness score; state-of-the-art VQA systems suffer large drops when forced to be both correct and consistent across paraphrases, while a cycle-consistency training scheme improves robustness without extra annotations. Distribution-shifted re-test sets—ImageNetV2 \cite{recht2019imagenetv2}—demonstrate a benign but consequential shift: across many architectures accuracy drops by roughly 11–14\% yet model rankings remain stable, challenging claims of test-set saturation. Complementing this, ObjectNet \cite{NIPS2019_9142} controls backgrounds, rotations, and viewpoints and yields about 40–45 percentage-point drops versus ImageNet, exposing spurious biases tied to scene context and pose. Synthesizing evidence across many real-world shifts, The Many Faces of Robustness \cite{hendrycks2021imagenet_r} introduce datasets such as ImageNet-R, SVSF, DFR, and Real Blurry Images, showing that scale and data augmentation can transfer to some shifts, but no single method uniformly helps, arguing for multi-faceted evaluations rather than a single robustness score. WILDS\cite{koh2021wilds} formalizes domain and subpopulation shifts across ten datasets, emphasizing worst-group metrics for deployment realism. In NLP, Yuan et al. \cite{yuan2023boss} propose BOSS (five tasks, twenty datasets) and find that many OOD methods bring limited gains over vanilla fine-tuning, while large language models with in-context learning can be preferable on OOD cases, again highlighting evaluation design over single-number reporting. Finally, LAION-C\cite{li2025laionc} argues ImageNet-C is often no longer truly OOD for web-scale models and releases a harder, automatically constructed corruption suite.

\begin{table*}[ht]
\centering
\scriptsize
\renewcommand{\arraystretch}{1.3}

\caption{
\textbf{Comprehensive survey of robustness benchmarks by category.}
This table summarizes representative benchmarks used to evaluate the robustness of AI models, organized by two top-level categories: adversarial robustness and natural robustness. It spans multiple robustness types, including adversarial, alignment, natural corruption, and out-of-distribution (OOD) generalization. For each benchmark, we report its publication year, venue, data size, robustness type, evaluation domain (abbreviated), and key evaluation metrics.
}

\resizebox{\textwidth}{!}{
\begin{tabular}{llllllll}
\toprule
\textbf{Category} & \textbf{Benchmark} & \textbf{Year} & \textbf{Venue} & \textbf{Data size} & \textbf{Robustness type} & \textbf{Evaluation domain} & \textbf{Metric(s)} \\
\midrule

\multirow{7}{*}{Adv} &
ANLI\cite{nie2020adversarial_nli} & 2020 & ACL & 3k test examples & Adversarial & Lang (NLI) & Accuracy \\
& \cellcolor{gray!15}AdvGLUE\cite{wang2021adversarial_glue} & \cellcolor{gray!15}2021 & \cellcolor{gray!15}NeurIPS & \cellcolor{gray!15}5k examples & \cellcolor{gray!15}Adversarial & \cellcolor{gray!15}Lang (GLUE) & \cellcolor{gray!15}Task-specific \\
& TruthfulQA\cite{lin2022truthfulqa} & 2021 & ACL & 817 questions & Alignment & Lang (QA) & Truthful response rate \\
& \cellcolor{gray!15}JailBreakV-28K\cite{luo2024jailbreakv} & \cellcolor{gray!15}2024 & \cellcolor{gray!15}COLM & \cellcolor{gray!15}28k adversarial cases & \cellcolor{gray!15}Adversarial & \cellcolor{gray!15}Multi & \cellcolor{gray!15}Attack success rate \\
& OODRobustBench\cite{li2024oodrobustbench} & 2024 & ICML & 23 natural shifts & Adversarial + OOD & Vision & Task-specific \\
& \cellcolor{gray!15}SEED‑Bench\cite{Li_2024_CVPR} & \cellcolor{gray!15}2024 & \cellcolor{gray!15}CVPR & \cellcolor{gray!15}19k questions & \cellcolor{gray!15}Alignment & \cellcolor{gray!15}Multi (MLLMs) & \cellcolor{gray!15}Accuracy \\
& AIR‑Bench\cite{yang2024airbench} & 2024 & ACL & 314 risk types & Alignment & Lang (LLMs) & Safe completion rate \\

\midrule

\multirow{9}{*}{Natural} &

ImageNet‑A\cite{hendrycks2019natural_adversarial} & 2019 & CVPR & 7.5k images & Natural adversarial & Vision & Accuracy \\
& \cellcolor{gray!15}ImageNet‑C\cite{hendrycks2019benchmarkingCorruptions} & \cellcolor{gray!15}2019 & \cellcolor{gray!15}ICLR & \cellcolor{gray!15}50k × 15 corruptions & \cellcolor{gray!15}Natural corruption & \cellcolor{gray!15}Vision & \cellcolor{gray!15}Accuracy, mCE \\
& VQA‑Rephrasings\cite{shah2019arxiv_vqa_rephrasings} & 2019 & CVPR & 40k × 3 questions & Natural variation & Multi (VQA) & Accuracy \\
& \cellcolor{gray!15}ImageNet‑V2\cite{recht2019imagenetv2} & \cellcolor{gray!15}2019 & \cellcolor{gray!15}ICML & \cellcolor{gray!15}10k images & \cellcolor{gray!15}OOD & \cellcolor{gray!15}Vision & \cellcolor{gray!15}Accuracy \\
& ObjectNet\cite{NIPS2019_9142} & 2019 & NeurIPS & 50k images & OOD & Vision & Accuracy \\
& \cellcolor{gray!15}ImageNet‑R\cite{hendrycks2021imagenet_r} & \cellcolor{gray!15}2021 & \cellcolor{gray!15}ICCV & \cellcolor{gray!15}30k images & \cellcolor{gray!15}OOD & \cellcolor{gray!15}Vision & \cellcolor{gray!15}Accuracy \\
& WILDS\cite{koh2021wilds} & 2021 & ICML & 10 datasets & OOD & Mixed & Task-specific \\
& \cellcolor{gray!15}BOSS\cite{yuan2023boss} & \cellcolor{gray!15}2023 & \cellcolor{gray!15}NeurIPS & \cellcolor{gray!15}20 datasets & \cellcolor{gray!15}OOD & \cellcolor{gray!15}Lang (NLP) & \cellcolor{gray!15}Task-specific \\
& LAION‑C\cite{li2025laionc} & 2024 & ICLR & 300k images & Natural corruption & Vision & Accuracy, mCE \\

\bottomrule
\end{tabular}
}
\label{tab:robustness_type}
\end{table*}

\subsubsection{Synthesis and governance implications}

Improved performance under one type of distribution shift does not reliably transfer to others. While data augmentation, domain generalization, and test-time adaptation each address specific robustness issues, no single method provides uniform protection across natural corruptions, subpopulation shifts, and multimodal misalignment. Consequently, robustness evaluation is inherently multi-dimensional. A single aggregate accuracy score cannot capture whether a system remains reliable under deployment-relevant changes.

Robustness governance therefore requires moving from one-time accuracy reporting to evidence that a model remains reliable under foreseeable deployment shifts. In practice, robustness benchmarks can inform industrial entry standards by specifying worst-group performance, corruption tolerance, and OOD degradation limits for high-risk applications. However, benchmark scores should be paired with post-deployment monitoring because real environments evolve through sensor changes, population shifts, language drift, and adversarial adaptation. Governance programs should therefore require model cards or system cards that disclose robustness assumptions, supported domains, known failure modes, and re-evaluation triggers when data distributions change.
\hypertarget{Hallucination}{}
\vspace{-0.5em}
\subsection{Hallucination} \label{sec:hallu}

\subsubsection{Problem definition}

Hallucination in LLMs refers to fluent but factually incorrect or fabricated outputs, representing a model-level failure of factuality or faithfulness; its consequences are particularly severe when such outputs are used in domains such as healthcare, law, and finance. NLP research~\cite{huang2025survey} typically distinguishes \fix{factuality hallucination, in which outputs contradict facts or external knowledge (\eg, false dates, entities, or citations), indicating poor grounding in truth, from faithfulness hallucination, in which outputs deviate from the input or instruction (\eg, ignoring queries, contradictions, flawed reasoning), reflecting misalignment with context or intent.}

\subsubsection{Existing methods}
Mitigating hallucinations in LLMs and MLLMs requires interventions across data, training, and inference stages. Several effective strategies have been developed to address both factual inconsistency and modality misalignment.

\begin{itemize}

\item \fix{\textbf{Data-level methods.}} Current solutions mitigate hallucination by improving semantic diversity, visual grounding, and sample structure. Balanced construction of positive-negative sample pairs, particularly via contrastive learning on hallucinated texts, enhances model robustness \cite{jiang2024hallucination}. Region-level and pixel-level annotations strengthen visual detail modeling and spatial alignment \cite{jain2024vcoder}. Feedback-augmented strategies, such as Silkie, VIGC, and Woodpecker, employ model reflection or external validators (\eg, GPT-4V) to refine vision-language consistency efficiently \cite{wang2024vigc}.

\item \fix{\textbf{Training-stage methods.}} Current methods focus on overcoming language dominance and enhancing visual-linguistic alignment. LLMs trained via MLE risk overconfident hallucinations due to fluency bias; semantic entropy and abstention training counter this by modeling uncertainty \cite{tjandra2024fine}. Perturbed input construction improves robustness against structure-sensitive errors \cite{ding2024hallu}, while multi-objective training (\eg, MOCHa) optimizes both fluency and factuality \cite{ben2023mitigating}. In MLLMs, techniques such as FERRET~\cite{you2023ferret}, VCoder~\cite{jain2024vcoder}, and GROUNDHOG~\cite{zhang2024groundhog} introduce dense visual encoding for fine-grained comprehension, while RAI-30k~\cite{chen2023mitigating} offer structured region-level supervision.

\item \fix{\textbf{Inference-stage methods.}} Hallucinations stem from visual memory decay, prior overreliance, and decoding bias. MEMVR~\cite{zou2024look} and DeCo~\cite{wang2024mllm} re-inject visual signals during generation to preserve factual grounding. Semantic entropy~\cite{farquhar2024detecting} and VL-Uncertainty~\cite{zhang2024vl} provide uncertainty-aware abstention mechanisms. Woodpecker~\cite{yin2024woodpecker} and OPERA~\cite{huang2024opera} validate image-text consistency post hoc or during decoding to suppress hallucinated content. Self-Refinement~\cite{jiang2024hallucination} and Thought Rollback~\cite{chen2024toward} offer plug-and-play reasoning corrections by prompting introspection and dynamic rerouting. Generation Constraint Scaling~\cite{kollias2024generation} and OpenCHAIR~\cite{ben2023mitigating} incorporate probabilistic control and token-level factuality metrics to constrain output drift. \fix{Visual Attention Reasoning further combines hierarchical visual search with self-verification to strengthen evidence grounding and correct reasoning errors through backtracking~\cite{cai2026visualattention}.}

\end{itemize}

\subsubsection{Benchmarks}

To support standardized evaluation and facilitate future research, \cref{tab:hallucination_llmbenchmarks} presents representative benchmarks commonly used to assess hallucination in LLMs and MLLMs.

Evaluation of hallucinations in LLMs generally focuses on two key types: factuality and faithfulness. Factual hallucinations occur when the generated content contradicts real-world knowledge, while faithfulness hallucinations emerge when the model deviates from the provided context. Generative benchmarks such as TruthfulQA~\cite{lin2022truthfulqa} and REALTIMEQA~\cite{kasai2023realtime} assess the truthfulness of answers to open-domain or time-sensitive questions, emphasizing the detection of factual errors. In contrast, discriminative benchmarks like HaluEval~\cite{li2023halueval} and FELM~\cite{zhao2023felm} assess the model's ability to detect hallucinations in existing texts, using tasks like classification and ranking. Recent efforts, such as HaluEval 2.0~\cite{li2024dawn} and FACTOR~\cite{muhlgay2024generating}, enable fine-grained analysis of hallucinations across diverse domains, focusing on evaluation metrics like accuracy, precision, and recall.

For MLLMs, hallucination evaluation becomes more complex due to the integration of visual context. Benchmarks like CHAIR~\cite{muhlgay2024generating} and POPE~\cite{li2023evaluating} focus on object hallucinations in image captioning and binary QA tasks. Newer benchmarks, such as CIEM~\cite{hu2023ciem} and HaELM~\cite{Guan_2024_CVPR}, enable large-scale evaluation of hallucinations by automating data generation and leveraging LLMs for assessment. More specialized benchmarks like AMBER~\cite{wang2023llm} and RAH-Bench~\cite{chen2023mitigating} offer fine-grained analysis by combining generative and discriminative scoring across object, attribute, and relation hallucinations. These benchmarks support a variety of evaluation formats, including multi-class tasks and binary classification, with metrics such as F1 score.

\begin{table*}[t]
\centering
\renewcommand{\arraystretch}{1.3}
\setlength{\tabcolsep}{5pt}
\caption{\textbf{Overview of representative benchmarks for evaluating hallucinations in LLMs and MLLMs.} The benchmarks are categorized by hallucination type---factuality (Fact) and faithfulness (Faith) for LLMs, and category (C), attribute (A), and relation (R) for MLLMs---and by evaluation type: generative (Gen) or discriminative (Dis). A diverse range of metrics is used for assessment.
}
\resizebox{\textwidth}{!}{
\begin{tabular}{p{1.2cm}lllllp{1.5cm}l}
\toprule
\textbf{Model} & \textbf{Benchmark} & \textbf{Year} & \textbf{Venue} & \textbf{Data size} & \textbf{Hallucination type} & \textbf{Evaluation type} & \textbf{Metric} \\
\midrule
\multirow{14}{*}{LLMs}
& TruthfulQA~\cite{lin2022truthfulqa}  & 2022 & ACL & 817 & Fact & Gen & LLM judge, human \\
& \cellcolor{gray!15}REALTIMEQA~\cite{kasai2023realtime}  & \cellcolor{gray!15}2023 & \cellcolor{gray!15}NeurIPS & \cellcolor{gray!15}- & \cellcolor{gray!15}Fact & \cellcolor{gray!15}Gen & \cellcolor{gray!15}Acc, EM, F1 \\
& HaluEval~\cite{li2023halueval}  & 2023 & EMNLP & 35000 & Faith & Dis & Acc \\
& \cellcolor{gray!15}FreshQA~\cite{vu2024freshllms} & \cellcolor{gray!15}2023 & \cellcolor{gray!15}EMNLP & \cellcolor{gray!15}600 & \cellcolor{gray!15}Fact  & \cellcolor{gray!15}Gen & \cellcolor{gray!15}Human \\
& FELM~\cite{zhao2023felm} & 2023 & NeurIPS & 3948 & Fact \& Faith & Dis & Balanced acc., F1 \\
& \cellcolor{gray!15}PhD~\cite{yang2023new} & \cellcolor{gray!15}2023 & \cellcolor{gray!15}EMNLP & \cellcolor{gray!15}300 & \cellcolor{gray!15}Faith & \cellcolor{gray!15}Dis & \cellcolor{gray!15}Pre., rec., F1 \\
& ScreenEval~\cite{lattimer2023fast} & 2023 & EMNLP & 52 & Faith & Dis & AUC \\
& FACTOR~\cite{muhlgay2024generating} & 2024 & EACL & 4030 & Fact & Dis & Likelihood \\
& \cellcolor{gray!15}BAMBOO~\cite{dong2024bamboo} & \cellcolor{gray!15}2024 & \cellcolor{gray!15}LREC-COLING & \cellcolor{gray!15}400 & \cellcolor{gray!15}Faith & \cellcolor{gray!15}Dis & \cellcolor{gray!15}Pre \& Rec \& F1 \\
& LSum~\cite{feng2024improving} & 2024 & EMNLP & 6166 & Faith & Dis & Balanced acc. \\
& \cellcolor{gray!15}SAC$^3$~\cite{zhang2023sac3} & \cellcolor{gray!15}2024 & \cellcolor{gray!15}EMNLP & \cellcolor{gray!15}250 & \cellcolor{gray!15}Faith & \cellcolor{gray!15}Dis & \cellcolor{gray!15}AUC \\
& HaluEval 2.0~\cite{li2024dawn} & 2024 & ACL & 8352 & Fact  & Gen & MiHR, MaHR \\
& \cellcolor{gray!15}HALoGEN~\cite{ravichander2025halogen} & \cellcolor{gray!15}2025 & \cellcolor{gray!15}ACL & \cellcolor{gray!15}10923 & \cellcolor{gray!15}Fact \& Faith & \cellcolor{gray!15}Gen & \cellcolor{gray!15}H-score, response rate, utility score \\
& \cellcolor{gray!15}HalluLens~\cite{bang2025hallulens} & \cellcolor{gray!15}2025 & \cellcolor{gray!15}ACL & \cellcolor{gray!15}- & \cellcolor{gray!15}Fact & \cellcolor{gray!15}Dis & \cellcolor{gray!15}Acc, F1 \\
\midrule
\multirow{9}{*}{MLLMs}
& CHAIR~\cite{rohrbach2018object} & 2018 & EMNLP & 5,000 & C & Gen & CHAIR \\
& \cellcolor{gray!15}POPE~\cite{li2023evaluating} & \cellcolor{gray!15}2023 & \cellcolor{gray!15}EMNLP & \cellcolor{gray!15}3,000 & \cellcolor{gray!15}C & \cellcolor{gray!15}Dis & \cellcolor{gray!15}Acc, pre., rec., F1 \\
& \cellcolor{gray!15}MMHal-Bench~\cite{sun2024aligning} & \cellcolor{gray!15}2023 & \cellcolor{gray!15}EMNLP & \cellcolor{gray!15}96 & \cellcolor{gray!15}C & \cellcolor{gray!15}Gen & \cellcolor{gray!15}LLM assessment \\
& HaELM~\cite{Guan_2024_CVPR} & 2023 & CVPR & 5,000 & - & Gen & LLM assessment \\
& \cellcolor{gray!15}MME~\cite{fu2025video} & \cellcolor{gray!15}2024 & \cellcolor{gray!15}CVPR & \cellcolor{gray!15}1,457 & \cellcolor{gray!15}C\&A\&R & \cellcolor{gray!15}Dis & \cellcolor{gray!15}Acc, score \\
& MMBench~\cite{liu2024mmbench} & 2024 & ECCV & 3,217 & C\&A\&R & Dis & Acc \\
& \cellcolor{gray!15}M-HalDetect~\cite{gunjal2024detecting} & \cellcolor{gray!15}2024 & \cellcolor{gray!15}AAAI & \cellcolor{gray!15}4,000 & \cellcolor{gray!15}C & \cellcolor{gray!15}Dis & \cellcolor{gray!15}Reward model score \\
& FGHE~\cite{wang2024mitigating} & 2024 & MMM & 200 & C\&A\&R & Dis & Acc, pre., rec., F1 \\
& \cellcolor{gray!15}GAVIE~\cite{liu2024mitigatinghallucinationlargemultimodal} & \cellcolor{gray!15}2024 & \cellcolor{gray!15}ICLR & \cellcolor{gray!15}1,000 & \cellcolor{gray!15}- & \cellcolor{gray!15}Gen & \cellcolor{gray!15}LLM assessment \\
\bottomrule
\end{tabular}
}
\label{tab:hallucination_llmbenchmarks}
\end{table*}

\subsubsection{Synthesis and governance implications}

Hallucination mitigation now spans the full model lifecycle, from data construction and training objectives to decoding control and post-hoc validation. Nevertheless, these methods remain constrained by a shared limitation: most rely on proxy signals for factuality, faithfulness, or visual grounding, whereas real-world errors often combine multiple failure types simultaneously. The central trade-off lies between informativeness and reliability. While conservative systems may reduce unsupported claims through abstention, excessive refusal can degrade usefulness in open-ended tasks.

Hallucination mitigation is therefore closely tied to governance challenges of reliability, transparency, and accountability. In deployment, hallucinated outputs may lead to misinformation, legal exposure, medical risk, or financial harm, especially when AI systems are integrated into decision-support workflows. Hallucination controls should not be treated only as model-quality improvements, but also as governance mechanisms that require uncertainty disclosure, retrieval-augmented verification, human-in-the-loop review, domain-specific reliability audits, and traceable evidence sources. For high-risk use cases, regulators and industry standards may require systems to document factuality evaluation results, abstention behavior, escalation policies, and post-deployment incident reports.
\hypertarget{Interpretability}{}
\subsection{Interpretability} \label{sec:interp}

\subsubsection{Problem definition}

With the rapid advancement of deep learning, black-box models have become dominant due to their strong performance, yet their opaque decision-making creates challenges in high-stakes fields like medical image analysis, underscoring the need for interpretability. Interpretability can be categorized into active and passive\cite{xu2025interpretability}. Active interpretability enhances transparency by designing inherently explainable architectures (\eg, decision trees\cite{custode2023evolutionary}, knowledge graphs\cite{gaur2022iseeq}, additive models\cite{agarwal2021neural}) or incorporating interpretable modules during training such as capsule networks\cite{sabour2017dynamic}, Memory Wrap\cite{la2023self}, and Stack-NMN\cite{hu2018explainable}. Passive interpretability, by contrast, applies post hoc analyses of model weights, outputs, or internal representations, using behavior-based\cite{covert2021explaining}, attribution-based\cite{shrikumar2017learning}, or concept-based methods\cite{burns2022discovering} to uncover patterns and improve transparency without modifying the original model.

\subsubsection{Existing methods}

Mechanistic interpretability \cite{nanda2023progress} is an emerging field of AI research that aims to understand the internal workings of neural networks. Rather than treating models as black boxes, this approach emphasizes analyzing the internal structure of models by examining components such as weights, neurons, layers and activations to derive meaningful explanations for model behavior \cite{bereska_mechanistic_2024}.
In general, this approach adopts a reverse-engineering methodology to identify functional roles of specific network components. Mechanistic interpretability plays a critical role not only in understanding model decisions but also in facilitating downstream applications \cite{zhang_cross-modal_2025}. These include model editing and intervention \cite{basu2024understanding}, the enhancement of compositional generalization capabilities \cite{zarei2024understanding}, and the identification and mitigation of spurious correlations \cite{balasubramanian2024decomposing}. We discuss mechanistic interpretability with two representative approaches, dictionary learning and attribution methods.

\vspace{-0.2em}

\begin{itemize}

\item \fix{\textbf{Dictionary learning.}}
A key challenge in interpretability is superposition, where neurons encode multiple unrelated features, obscuring the meaning of individual activations\cite{elhage2022toy}. This entanglement is especially pronounced in deep networks with high-dimensional, distributed representations. Dictionary learning addresses this by decomposing activations into sparse combinations of simpler features\cite{lin2019sparse}, based on the hypothesis that disentangled components better capture semantic structure. Sparse Autoencoders\cite{huben2023sparse} are widely used for this purpose, reconstructing activations while enforcing sparsity to learn latent features aligned with human-understandable concepts. However, learned dictionaries can be unstable across runs\cite{rana2024interpretable}, individual atoms may lack clear semantics without human validation, and features risk capturing spurious correlations rather than causal mechanisms\cite{dictionary_learning_features_classifier}, limiting their reliability for interpretability and auditing. Recent mechanistic interpretability research further extends sparse-feature analysis toward larger language models through large-scale feature extraction, cross-layer transcoders, and circuit-tracing methods that attempt to connect learned features into functional computational pathways across model components\cite{brunello2025blackboxnlp}. These developments are relevant to governance because they may support more targeted audits of unsafe concepts, deceptive behavior, and spurious reasoning patterns. Their use as governance evidence, however, still depends on feature validation, causal testing, and scalability across model families and deployment settings.

\item \fix{\textbf{Attribution.}}
Attribution methods aim to explain model behavior by assigning responsibility to input features, internal components (\eg, attention heads, neurons, layers), or training examples. These post hoc tools analyze predictions and activations to enhance transparency. Gradient-based methods such as Integrated Gradients\cite{sundararajan2017axiomatic}, Grad-CAM\cite{selvaraju2017grad}, SmoothGrad\cite{smilkov2017smoothgrad}, and DeepLIFT\cite{shrikumar2017learning} estimate how small input perturbations affect outputs, revealing input–output relationships without accessing internal structures. Beyond input-level attribution, techniques like Direct Logit Attribution quantify the influence of specific neurons on predictions for finer-grained insights\cite{dar2023analyzing}, though both approaches often capture correlations rather than causality. Data attribution complements these methods by tracing outputs to influential training instances using techniques such as influence functions\cite{koh2017understanding}. While attribution methods struggle with out-of-distribution data and emergent misaligned behaviors, combining model- and data-centric perspectives provides a richer understanding of predictions and their underlying drivers.

\end{itemize}

\subsubsection{Governance implications}
Interpretability connects technical transparency with institutional accountability. For governance purposes, explanation tools should support auditability, contestability, and traceability rather than merely producing visual saliency maps or post hoc narratives. In high-impact domains, interpretable evidence can help regulators and auditors verify whether a system relies on prohibited attributes, spurious correlations, or unstable reasoning paths. Practical governance mechanisms may include explanation documentation in model cards, mandatory interpretability reports for high-risk systems, reproducible audit logs, and requirements that explanations be understandable to the relevant stakeholder group, such as clinicians, compliance officers, affected users, or independent auditors.
\hypertarget{Derivative security}{}
\section{Derivative security} \label{sec:derivative}

Derivative security refers to externally manifested risks that arise when AI models, data, outputs, users, and deployment contexts interact. These risks may originate from training data, learned representations, model parameters, or generated outputs, but they become governance concerns through their effects on external parties, such as privacy leakage, discriminatory treatment, or large-scale misuse. In this section, we focus on three aspects: data privacy, bias and discrimination, and abuse and misuse of AI.

\hypertarget{Privacy risks and mitigation}{}
\subsection{Data privacy} \label{sec:privacy}

\subsubsection{Privacy attacks}
Privacy attacks seek to extract, infer, or reconstruct private or confidential information from LLMs by exploiting their outputs, interaction histories, learned representations, or deployment interfaces. Such information may originate from training data, user queries, contextual memory, or proprietary model assets, and the resulting risk is an unintended information flow from protected parties to unauthorized observers. Major threats include interactive data leakage, inference-based attacks, and deployment-time exploits.

\noindent\fix{\textbf{(1) Data leakage threats.}} LLMs, trained on vast datasets and processing user queries, can leak private data from outputs or internal states via well-designed prompting.

\begin{itemize}
    \item \fix{\textbf{Sensitive query leakage.}} LLMs may inadvertently reveal private data from user prompts or conversation history if specific details are ``remembered" and included in outputs where they shouldn't be recalled \cite{iqbal2024llm}.

    \item \fix{\textbf{Contextual leakage.}} Accumulating seemingly harmless details over time from the context of usage, such as conversation history or integrated data sources, can lead to the inference of private information\cite{staab2023beyond}.

    \item \fix{\textbf{Personal-preference leakage.}} LLMs can inadvertently reveal or allow inference of a user's personal attributes, behaviors, or preferences through their responses, even accurately inferring details like location or income~\cite{harte2023leveraging}.
\end{itemize}

\noindent\fix{\textbf{(2) Inference-time threats.}} These threats exploit a deployed LLM's output behavior to infer unauthorized information about its training data or properties.

\begin{itemize}
    \item \fix{\textbf{Membership inference attacks (MIAs).}} MIAs aim to determine if specific data was part of an LLM's training set. While challenging for large, generalized LLMs, advanced techniques leveraging likelihood ratios \cite{mireshghallah2022quantifying}, synthetic neighbors \cite{mattern2023membership}, or self-prompt calibration \cite{fu2024membership} show improved effectiveness, especially against fine-tuned models. Even ``label-only'' attacks \cite{he2025towards}, accessing only generated text, can be effective against pre-trained LLMs.
    \item \fix{\textbf{Attribute inference attacks.}} These attacks infer sensitive attributes (\eg, location, income, health) about individuals whose data is used for training or interaction, based on the LLM's output or internal representations \cite{pan2020privacy}.
\end{itemize}

\noindent\fix{\textbf{(3) Deployment-time threats.}} These attacks target the deployed LLM or its infrastructure to extract parameters, manipulate behavior, or infer model properties.

\begin{itemize}
    \item \fix{\textbf{Model inversion attacks.}} Studies show that verbatim training data, including personally identifiable information, can be extracted from LLMs, with larger models being more vulnerable \cite{carlini2021extracting}.
    \item \fix{\textbf{Model stealing attacks.}} Adversaries with query access can reconstruct proprietary LLMs through query-based extraction, potentially exposing sensitive data or intellectual property—even with limited queries or no access to original training data \cite{truong2021data}.
    \item \fix{\textbf{Backdoor attacks.}} Malicious functionality is injected during training, causing normal behavior on clean inputs but attacker-defined outputs when a specific ``trigger'' is present. These can manipulate LLM responses to be biased, harmful, or privacy-violating through data poisoning~\cite{wan2023poisoning}, weight modification~\cite{huang2024composite}, or instruction tuning \cite{xu2024instructions}.
\end{itemize}

\subsubsection{Privacy defenses}

Addressing LLM privacy threats requires a multi-faceted approach throughout the model's life-cycle.

\noindent\fix{\textbf{(1) Training-time defenses.}} These defenses are applied during LLM creation, embedding privacy protection directly into the model's learning process or training data.

\begin{itemize}
    \item \fix{\textbf{Data-oriented defenses.}} These focus on preprocessing or modifying training data to reduce privacy risks. Deduplicating training data significantly enhances security against data extraction and memorization attacks \cite{kandpal2022deduplicating}. Detecting personal information in corpora is also crucial, though current methods have limitations \cite{subramani2023detecting}.
    \item \fix{\textbf{Differential privacy-based training.}} Differential Privacy (DP) protects against individual data leakage by adding noise during training (\eg, to gradients). Recent advances mitigate performance drops by using large pre-trained models and fine-tuning. Methods like private word-piece algorithms \cite{hoory2021learning}, EW-Tune \cite{behnia2022ew}, and DP for Parameter-Efficient Fine-Tuning (PEFT) \cite{li2023privacy} improve model utility. Other approaches, such as DP-Forward \cite{du2023dp}, Adaptive DP \cite{wu2022adaptive}, and Selective DP \cite{shi2022just}, further optimize privacy-utility trade-offs. DP has also been explored for enforcing the ``Right to be Forgotten'' in LLMs \cite{zhang2024right}.
    \item \fix{\textbf{Knowledge unlearning.}} Knowledge Unlearning efficiently removes the influence of specific data from trained models, which is crucial for ``Right to be Forgotten''. The challenge, however, is how to conduct unlearning from massive LLMs without full retraining. Efficient frameworks use lightweight unlearning layers \cite{chen2023unlearn} or approximate techniques, such as identifying related tokens, to erase content with minimal performance impact \cite{eldan2023s}.
\end{itemize}

\noindent\fix{\textbf{(2) Inference-time defenses.}} These defenses are applied during LLM use, protecting user queries, the model's internal state and generated output from privacy breaches.
\begin{itemize}
    \item \fix{\textbf{Secure computation-based defenses.}} Secure Multi-Party Computation (MPC) and Function Secret Sharing (FSS) enable joint computation over private inputs without revealing them. These techniques allow privacy-preserving LLM inference, protecting both user prompts and model parameters. Advances in secure matrix multiplication \cite{hou2023ciphergpt}, GELU \cite{ding2023east}, and Softmax \cite{liu2023llms} for GPT inference, along with frameworks like PUMA \cite{dong2023puma}, enable efficient secure inference for large models.
    Confidential Computing, leveraging hardware Trusted Execution Environments (TEEs), offers another approach by creating secure enclaves for data and computation, even from cloud providers \cite{chen2023verified}. Combining PEFT with distributed privacy-sensitive computation also offers efficient LLM services \cite{wang2023privatelora}.
    \item \fix{\textbf{Detection-based defenses.}} These defenses monitor LLM interactions to identify patterns or outputs indicating a potential privacy breach or attack \cite{mireshghallah2023can}. However, specific technical mechanisms for detecting privacy violations during inference are still an open challenge.
\end{itemize}

Table~\ref{tab:privacy_benchmarks_concise} summarizes representative attack-, protection-, and evaluation-oriented privacy benchmarks and their principal metrics.

\begin{table*}[t]
\renewcommand{\arraystretch}{1.3}
\caption{\textbf{Overview of representative privacy benchmarks for LLMs.} This table summarizes key benchmarks and evaluation frameworks for language-model privacy research. Atk denotes attack-oriented evaluations, Pro denotes protection- or defense-oriented evaluations, and Eval denotes broader privacy-risk evaluations.}
\centering
\resizebox{\textwidth}{!}{
\begin{NiceTabular}{llll}
\CodeBefore
    \rowcolors{2}{gray!15}{}
\Body
\toprule
\textbf{Category} & \textbf{Benchmark} & \textbf{Keyword} & \textbf{Key metrics} \\
\midrule
\Block[fill=white]{5-1}{\makecell{Privacy\\Attack}}
& ArxivMIA~\cite{liu2024probing} & Atk-MIA & AUC, TPR@FPR \\
& WikiMIA~\cite{shi2023detecting} & Atk-MIA & AUC, TPR@FPR \\
& MIMIR~\cite{duan2024membership} & Atk-MIA & AUC, TPR@low\%FPR \\
& ProPILE~\cite{kim2023propile} & Atk-PII leakage & String match, likelihood \\
\midrule
\Block[fill=white]{5-1}{\makecell{Privacy\\Evaluation\\\& Protection}}
& TOFU~\cite{maini2024tofu} & Pro-Knowledge unlearning & Forget quality (KS test), model utility \\
& PrivLM-Bench~\cite{li2023privlm} & Privacy evaluation & AUC, TPR, exposure, micro-F1 \\
& CONFAIDE~\cite{mireshghallah2023can} & Pro-Contextual inference & Pearson correlation, leakage rate \\
& LLM-PBE~\cite{li2024llmpbe} & Eval-lifecycle privacy & Extraction acc., MIA AUC, TPR@FPR \\
& PrivacyLens~\cite{shao2024privacylens} & Eval-Contextual privacy & Accuracy, leakage rate \\
\bottomrule
\end{NiceTabular}
}
\label{tab:privacy_benchmarks_concise}
\end{table*}

\subsubsection{Synthesis and governance implications}

Despite significant advances along two complementary axes---training-time protections such as differential privacy and unlearning, and inference-time protections such as secure computation and leakage detection---privacy evaluation remains fragmented. Current benchmarks often assess these mechanisms in isolation, while real deployments combine training data, user prompts, retrieval components, model updates, and service infrastructures. Privacy thus extends beyond preventing a single leakage event; it demands lifecycle governance over how information is memorized, inferred, transmitted, or reconstructed over time.

Privacy governance must therefore navigate the tension between large-scale data utilization and responsible AI governance. Techniques such as differential privacy, secure computation, unlearning, and data filtering provide important safeguards, but they must be embedded within legal and organizational controls over consent, data ownership, retention, access, and accountability. From a governance perspective, privacy protection should be supported by continuous privacy audits, data lineage records, risk-based access controls, and compliance-aware training pipelines. These mechanisms help translate technical privacy defenses into enforceable obligations under data-protection regimes and sector-specific standards.

\hypertarget{Bias and discrimination}{}
\subsection{Bias and discrimination} \label{sec:bias}

\subsubsection{Bias/discrimination attacks}

Bias and discrimination attacks aim to exploit or amplify unequal treatment in AI systems, turning biased data, learned representations, model objectives, or interaction patterns into discriminatory outcomes for external individuals or groups. These attacks can be broadly organized into three layers: data-layer manipulation, model- or algorithm-level exploitation, and interaction-layer hijacking.

\noindent\fix{\textbf{(1) Data-layer manipulation.}} At the data layer, biased or manipulated training data can encode discriminatory patterns into downstream models. A widely discussed example is Amazon's experimental recruiting tool, which reportedly learned gender-biased patterns from historical resumes and penalized resumes containing terms such as ``women's''~\cite{dastin2018amazon}. Beyond naturally occurring dataset bias, training data can also be deliberately poisoned to worsen group fairness or manipulate fairness-related measures~\cite{solans2021poisoning,mehrabi2021exacerbating}.

\noindent\fix{\textbf{(2) Model-level exploitation.}} At the model level, biased behavior may result from learned representations, optimization objectives, or hidden conditional mechanisms. Backdoor and Trojan attacks show that models can behave normally on clean inputs while producing attacker-specified outputs when triggers appear~\cite{gu2017badnets}. In fairness-sensitive settings, such mechanisms are concerning when triggers correlate with demographic or contextual cues. Empirical studies also show that vision systems can exhibit unequal performance across demographic groups. For example, commercial gender classification systems have shown large intersectional accuracy disparities~\cite{buolamwini2018gender}, and object detectors have exhibited unequal predictive performance across pedestrian skin-tone groups in driving-scene data~\cite{wilson2019predictive}. These findings indicate demographic performance gaps in perception modules, rather than intentional vehicle-level prioritization.

\noindent\fix{\textbf{(3) Interaction-layer hijacking.}} At the interaction layer, prompts, role-playing, or jailbreaks may expose or amplify biased behavior in LLMs. Bias benchmarks show that language models can rely on stereotypes in ambiguous question-answering contexts~\cite{parrish2022bbq}, while jailbreak studies show that safety-trained LLMs may still be induced to produce undesired outputs under adversarial prompts~\cite{wei2023jailbroken}. In sensitive settings such as hiring, lending, healthcare, and legal decision support, such failures may translate latent biases into discriminatory outcomes.

\subsubsection{Bias/discrimination defenses}
Addressing bias in AI and LLMs requires a multifaceted approach that spans the entire life-cycle of model development and deployment. A primary strategy involves using diverse and representative training data that accurately reflects the population the model is intended to serve. Employing various bias detection and debiasing tools and algorithms is also crucial for identifying and rectifying biases in both the data and the models~\cite{mehrabi2021survey}. Continuous monitoring of AI systems after deployment is crucial for detecting any emerging biases or performance shifts across different demographic groups. In critical decision-making areas where AI biases could have profound ethical or legal implications, incorporating human oversight is a vital safeguard.

Additional technical mitigation strategies include fairness-aware training, which explicitly optimizes fairness metrics~\cite{zafar2017fairness}, as well as data augmentation methods that ensure the balanced representation of various demographic groups. Prompt engineering can also help reveal and mitigate biases in LLMs, while fine-tuning models with debiasing objectives or datasets is another effective approach for reducing bias over time~\cite{cheng2024rlrf}.

\subsubsection{Benchmarks}

To facilitate standardized evaluation and future research, Table~\ref{tab:bias_benchmarks} presents representative benchmarks for evaluating bias and discrimination in NLP and CV domains.

In NLP, various tasks such as sentiment analysis, machine translation, and question answering (QA) are adopted to evaluate bias. For sentiment analysis, frameworks like Bias-BERT~\cite{venugopal2024comprehensive} and CALM~\cite{gupta2023calm} assess demographic bias, particularly concerning gender and race, using metrics such as F1 score. In machine translation, gender bias has been observed in systems like Google Translate, leading to the development of benchmarks such as MT-GenEval~\cite{currey2022mt}, which evaluates gender translation accuracy across languages. The BBQ dataset~\cite{parrish2021bbq} assesses social bias in QA models, focusing on biases against protected classes, while KoBBQ~\cite{jin2024kobbq} adapts this to the Korean context. Other QA benchmarks, like NovelQA~\cite{wang2024novelqa} and MEQA~\cite{li2024meqa}, evaluate the fairness of models in complex multi-hop and extended narrative tasks. For text generation, FairPrism~\cite{fleisig2023fairprism} targets fairness in generated text, addressing gender and sexual biases.

In CV, bias often manifests in image classification and facial recognition tasks. Datasets like FewSTAB~\cite{zheng2024benchmarking} evaluate how few-shot image classifiers perform across different demographic groups, while VLBiasBench~\cite{wang2024vlbiasbench} targets biases in large vision-language models. The FACET~\cite{gustafson2023facet} benchmark assesses fairness in image classification, object detection, and segmentation, highlighting the need for fairness-aware training. Additionally, the Fair SA framework~\cite {joshi2022fair} measures group fairness in face recognition, identifying disparities related to gender and skin tone. These efforts emphasize the importance of using specialized benchmarks to mitigate bias in CV systems and ensure equitable outcomes.

\begin{table*}[t]
\centering
\renewcommand{\arraystretch}{1.3}
\caption{\textbf{Representative benchmarks for bias and discrimination evaluation.} This table summarizes key benchmarks for bias and discrimination evaluation in natural language processing (NLP) and computer vision (CV) tasks. The NLP tasks include machine translation (MT), question answering (QA), sentiment analysis (SA), natural language inference (NLI), and text generation (TG). The CV tasks include image classification, facial recognition, and visual question answering.}
\resizebox{\textwidth}{!}{
\begin{tabular}{lllllll}
\toprule
\textbf{Domain} & \textbf{Benchmark} & \textbf{Year} & \textbf{Venue} & \textbf{Data size} & \textbf{Task} & \textbf{Metric} \\
\midrule
\multirow{7}{*}{NLP}
 & MT-GenEval~\cite{currey2022mt} & 2022 & EMNLP & 4K & MT & Acc, gender quality gap \\
 & \cellcolor{gray!15}BBQ~~\cite{parrish2021bbq} & \cellcolor{gray!15}2022 & \cellcolor{gray!15}ACL & \cellcolor{gray!15}58K & \cellcolor{gray!15}QA & \cellcolor{gray!15}Acc, bias score \\
 & \cellcolor{gray!15}FairPrism~\cite{fleisig2023fairprism} & \cellcolor{gray!15}2023 & \cellcolor{gray!15}ACL & \cellcolor{gray!15}5K & \cellcolor{gray!15}TG & \cellcolor{gray!15}Bias type annotation, harm extent \\
 & Stowe et al.\cite{stowe2024identifying} & 2024 & BEA & 601 & TG & F1 \\
 & \cellcolor{gray!15}KoBBQ~~\cite{jin2024kobbq} & \cellcolor{gray!15}2024 & \cellcolor{gray!15}TACL & \cellcolor{gray!15}76K & \cellcolor{gray!15}QA & \cellcolor{gray!15}Acc, bias score \\
 & MEQA~\cite{li2024meqa} & 2024 & NeurIPS & 2K & QA (multi-hop event-centric) & F1, pre., rec., completeness, logical consistency \\
 & \cellcolor{gray!15}NovelQA~\cite{wang2024novelqa} & \cellcolor{gray!15}2025 & \cellcolor{gray!15}ICLR & \cellcolor{gray!15}2K & \cellcolor{gray!15}QA (extended narratives) & \cellcolor{gray!15}Acc \\
 & FairMT-Bench~\cite{fan2024fairmt} & 2025 & ICLR & 10K & TG (multi-turn dialogue) & Bias ratio \\
\midrule
\multirow{3}{*}{CV}
 & \cellcolor{gray!15}Fair SA~\cite {joshi2022fair} & \cellcolor{gray!15}2022 & \cellcolor{gray!15}AFCR & \cellcolor{gray!15}200K & \cellcolor{gray!15}Facial recognition & \cellcolor{gray!15}AUC \\
 & FACET~\cite{gustafson2023facet} & 2023 & ICCV & 32K & Image classification, object & Acc, rec. \\ & & & & & detection, and segmentation & \\
 & \cellcolor{gray!15}FewSTAB~\cite{zheng2024benchmarking} & \cellcolor{gray!15}2024 & \cellcolor{gray!15}ECCV & \cellcolor{gray!15}850K & \cellcolor{gray!15}Image classification & \cellcolor{gray!15}Acc, worst accuracy (wAcc) \\
\bottomrule
\end{tabular}
}
\label{tab:bias_benchmarks}
\end{table*}

\subsubsection{Synthesis and governance implications}

While bias research has progressed from isolated detection toward pipeline-level mitigation, current evaluation remains fragmented. Many benchmarks focus on a restricted set of demographic categories, tasks, languages, or modalities, obscuring intersectional and context-specific harms. More importantly, fairness interventions involve normative choices: reducing disparity for one group or metric may leave other forms of inequity unresolved, and aggregate performance often masks subgroup failures. Bias governance therefore cannot be inferred from a single fairness score.

Bias and fairness issues extend beyond technical model behavior because they directly affect social trust, inclusion, and institutional legitimacy. Fairness evaluation should therefore be treated as a governance mechanism for accountability and social responsibility, not merely as a statistical optimization objective. Future governance frameworks may require demographic transparency reports, fairness audits, third-party evaluation, human appeal channels, and continuous monitoring of deployed outcomes across protected and context-specific groups. Because fairness definitions vary across legal, cultural, and sectoral contexts, governance-aware evaluation should make the chosen fairness criteria explicit and justify them for the relevant deployment setting.

\subsection{Abuse and misuse} \label{sec:abuse}
\subsubsection{Deepfake attacks}
Deepfake generation is enabled by advanced generative models, including Generative Adversarial Networks (GANs), Variational Autoencoders (VAEs), and diffusion models for visual and audio synthesis, as well as LLMs for text generation. These technologies produce highly realistic synthetic content across various modalities, including face-swapped videos, cloned voices, and generated text. While supporting creative applications, they also pose risks when used for impersonation or the dissemination of false information. This section reviews key generation techniques and their potential for misuse.

\noindent \fix{\textbf{(1) Text-based methods.}} Advanced LLMs like GPT-4o~\cite{openai_gpt4o_2024} and Gemini 2.5 ~\cite{google_gemini_2_5_2025} can generate fluent and context-aware text. Open-source models such as DeepSeek-V3~\cite{deepseekai2025deepseekv3technicalreport} and Qwen2.5~\cite{qwen2025qwen25technicalreport} further reduce the entry barrier. Although these tools offer many benefits, they can also be misused for tasks such as producing fake news, impersonating individuals, and launching phishing attacks.

\noindent \fix{\textbf{(2) Image/video deepfake generation.}} Recent advances in GANs~\cite{goodfellow2014generative} and diffusion models~\cite{ho2020denoising} enable the realistic generation of facial images and videos. These techniques lower the barrier for creating visual deepfakes, raising concerns about misuse, such as spreading false information.

\begin{itemize}
    \item \fix{\textbf{Face swapping.}} Replaces a person’s face in an image or video with another’s while maintaining pose and expression~\cite{Zhao_2023_CVPR}. This can be used to fabricate visual identities and mislead viewers.

    \item \fix{\textbf{Facial attribute editing.}} Alters facial attributes such as age, gender, or expression~\cite{kumar2024efficient3dawarefacialimage}, enabling subtle manipulations for identity concealment or deceptive narratives.

    \item \fix{\textbf{Face reenactment.}} Transfers facial motion or expression from a source to a target~\cite{rochow2024fsrt}, allowing realistic imitation of actions or emotions not performed by the individual.

    \item \fix{\textbf{Talking-face generation.}} Synthesizes speech-aligned facial movements from audio or text~\cite{Prajwal_2020_CVPR, Zhao_2025_CVPR}, enabling the creation of highly realistic talking-head videos for fake public communication or impersonation.
\end{itemize}

\fix{These techniques lower the barrier for visual forgery and can be exploited for large-scale impersonation, disinformation, and reputational harm, raising critical challenges for detection and governance.}

\noindent \fix{\textbf{(3) Audio deepfakes.}}
Audio deepfakes synthesize human-like speech from text or acoustic inputs using neural models. Advances in generative architectures—including GAN-based methods~\cite{kumar2019melgan} and VAE-based methods~\cite{lu2021vaenar}, diffusion-based methods~\cite{huang2022fastdiff}, and transformer-based methods~\cite{ren2019fastspeech}—have significantly enhanced the quality of speech synthesis. While these technologies enable beneficial applications like personalized voice assistants and accessibility tools, they also raise ethical concerns by enabling voice impersonation, phone scams, and audio disinformation.

\noindent \fix{\textbf{(4) Multimodal deepfakes.}}
Multimodal deepfakes refer to the generation of synthetic content across multiple modalities, such as images, text, audio, and video, using large generative models. Recent models, including AnyGPT~\cite{zhan2024anygpt}, NExT-GPT~\cite{wu2024next}, AudioGPT~\cite{huang2024audiogpt}, and Google’s Veo3~\cite{google2025veo3}, enable the creation of synthetic content across multiple modalities while ensuring a high level of consistency.

These advances make it easier to create realistic, multi-modal simulations of people or events, raising new concerns about coordinated manipulation, fabricated identities, and cross-media misinformation~\cite{Lin2024DetectingMG}. Addressing these challenges requires safeguards that can keep pace with the growing complexity of multimodal content.

\subsubsection{Deepfake defenses}
Detecting synthetic content is essential for preventing the harmful use of generative models. As deepfakes become more realistic and widely available, strong detection methods are needed to spot fake text, audio, and video. These methods help reduce risks such as impersonation and deception. This section introduces the primary detection approaches and explains how they support content security and compliance with regulations.

\noindent \fix{\textbf{(1) Text deepfake detection.}} Detection methods for text-based deepfakes are typically classified as white-box or black-box, depending on whether the internal structure of the generation model is accessible~\cite{wu2025survey}.

\begin{itemize}
    \item \fix{\textbf{White-box methods.}}
    These methods leverage internal model signals, such as token probabilities or hidden states. A representative strategy is statistical watermarking~\cite{kirchenbauer2023watermark}, which embeds imperceptible patterns during generation for later attribution. While effective under controlled conditions, these methods require access to the model internals and are less applicable in open or adversarial settings.
    \item \fix{\textbf{Black-box methods.} These methods analyze only generated text without requiring access to the underlying model, making them suitable for model-agnostic and practical deployment. Common approaches include \textit{zero-shot perturbation analysis}, which estimates the likelihood of generation by measuring semantic consistency under input variations~\cite{mitchell2023detectgpt}; \textit{fine-tuned classifiers}, which are trained on labeled datasets to distinguish between human-written and machine-generated text~\cite{liu2023argugpt}; \textit{adversarial training}, which improves robustness against adaptive synthetic content~\cite{DBLP:conf/nips/HuCH23}; and \textit{LLMs-as-detectors}, which prompt LLMs to evaluate the authenticity of a given input~\cite{lucas2023fighting}.}
\end{itemize}

\noindent \fix{\textbf{(2) Image/video deepfake detection.}} Deepfake detection in images and videos focuses on forgery cues in the spatial, frequency, and temporal domains. In parallel, data-driven methods learn to recognize manipulation patterns directly from large datasets. Together, these approaches support content verification and help counter visual misinformation.
\begin{itemize}
    \item \fix{\textbf{Spatial-domain methods.}} These approaches detect visual artifacts within individual frames, such as blending boundaries, unnatural textures, or inconsistent facial attributes. Representative strategies include artifact localization through reconstruction errors~\cite{Cao_2022_CVPR_RECCE}, and gradient-based feature encoding for cross-dataset generalization~\cite{tan2023learning_LGrad}.
    \item \fix{\textbf{Frequency-domain methods.}} These methods operate in the spectral domain, modeling anomalies in frequency components, compression patterns, or phase-amplitude mismatches. For instance, F$^3$-Net~\cite{qian2020thinking_F3Net} leverages frequency decomposition, while FreqNet~\cite{tan2024frequencyawareFreqNet} captures source-agnostic spectral signatures to improve robustness.
    \item \fix{\textbf{Temporal-domain methods.}} Temporal methods capture inter-frame inconsistencies such as lip-sync errors. LipForensics~\cite{haliassos2021lips_LipForen} models audio-visual coherence, and TI$^2$Net~\cite{Liu_2023_WACV_TI2Net} introduces identity-aware contrastive modeling for tracking semantic consistency across frames.
\end{itemize}
Although detection methods have advanced, they often lack robustness against unseen deepfake techniques, generalizing a central challenge for reliable content authentication and a pressing concern in long-term media trust.

\noindent \fix{\textbf{(3) Audio-based deepfake detection.}} Audio deepfakes involve the manipulation or generation of speech-like audio, including synthesized voices, tampered recordings, or cross-lingual voice conversion. Detection methods mainly rely on deep learning models that process raw waveforms or spectrograms to extract spectral and temporal artifacts. Representative approaches include CNN-based models, such as AASIST~\cite{jung2022aasist}, raw waveform classifiers like RawNet2~\cite{tak2021end}, graph-based networks that model spectral-temporal dependencies~\cite{tak2021end1}, and transformer-based hybrids that combine convolution and attention~\cite{bartusiak2022transformer}. In multimodal scenarios, fusion models such as FRADE and AVFakeNet~\cite{ILYAS2023110124} exploit cross-modal cues to improve reliability.

From a governance standpoint, these detection methods help prevent the misuse of synthetic audio in fraud and deception, supporting trust in voice systems and encouraging responsible use of speech technologies.

\noindent \fix{\textbf{(4) Multimodal deepfake detection.}} Multimodal deepfake detection addresses forged content spanning text, audio, image, and video by analyzing semantic or temporal inconsistencies between modalities. Compared to unimodal detection, this task is more complex due to the need for cross-modal coherence. Recent works explore visual-text alignment~\cite{shao2023dgm4}, audiovisual synchronization~\cite{kharel2023df_DF_Transfusion}, and zero-shot detection~\cite{ren2025can} using multimodal large language models. As generative models evolve to integrate multiple modalities, robust multimodal detection becomes crucial for preserving information integrity.

To support standardized evaluation and facilitate future research,~\cref{tab:deepfake_benchmarks} presents benchmarks for deepfake generation and detection, respectively, which cover data modalities and scales, and commonly used evaluation metrics.

\begin{table*}[t]
\renewcommand{\arraystretch}{1.3}
\caption{\textbf{Representative benchmarks for deepfake generation and detection.} This table summarizes key benchmarks for deepfake generation and detection. Metrics include Fréchet inception distance (FID), peak signal-to-noise ratio (PSNR), Fréchet video distance (FVD), multiscale structural similarity (MS-SSIM), kernel inception distance (KID), facial action unit (AU), accuracy (ACC), area under the curve (AUC), and equal error rate (EER).}
\centering
\resizebox{\textwidth}{!}{
\begin{NiceTabular}{lllllll}
\CodeBefore
    \rowcolors{2}{gray!15}{}
\Body
\toprule
\textbf{Task} & \textbf{Benchmark} & \textbf{Year} & \textbf{Venue} & \textbf{Data size} & \textbf{Type} & \textbf{Metric} \\
\midrule
\Block[fill=white]{9-1}{\makecell{Deepfake\\generation}}
 & FFHQ~\cite{karras2019style} & 2019 & CVPR & 70K & Image & FID, AU \\
 & CelebAMask-HQ~\cite{lee2020maskgan} & 2020 & CVPR & 30K & Image & FID \\
 & MEAD~\cite{wang2020mead} & 2020 & ECCV & 281K & Video & FID \\
 & CelebV-HQ~\cite{zhu2022celebv} & 2022 & ECCV & 30K & Video & FVD, FID \\
 & Husseini et al.~\cite{husseini2023comprehensive} & 2023 & ICCV & 240 & Video & SSIM, CSIM, LPIPS, AKD, FID, FVD\\
 & VBench~\cite{huang2024vbench} & 2024 & CVPR & 1.6K & Video & RAFT and 15 other metrics \\
 & EFHQ~\cite{dao2024efhq} & 2024 & CVPR & 450K & Multimodal & FIT, AKD, AED\\
 & AI-Face~\cite{lin2025ai} & 2025 & CVPR & 1600K & Image & FID, KID\\
 & DualTalk~\cite{peng2025dualtalk} & 2025 & CVPR & 5K & Video & FD, PFD, MSE, SID, rPCC\\
\midrule
\Block[fill=white]{13-1}{\makecell{Deepfake\\\fix{detection}}}
 & FaceForensics++~\cite{rossler2019faceforensics++} & 2019 & ICCV & 6K & Video & ACC, AUC \\
 & Celeb-DF~\cite{li2020celeb} & 2019 & CVPR & 6K & Video & ACC, AUC \\
 & DFFD~\cite{stehouwer2019detection} & 2020 & CVPR & 300K & Image & ACC, AUC \\
 & Deeperforensics~\cite{jiang2020deeperforensics} & 2020 & CVPR & 60K & Video & ACC, AUC \\
 & ForgeryNet~\cite{he2021forgerynet} & 2021 & CVPR & 2900K & Image & ACC, AUC \\
 & FakeAVCeleb~\cite{khalid2021fakeavceleb} & 2021 & NeurIPS & 20K & Multimodal & ACC, AUC \\
 & ADD 2022~\cite{yi2022add} & 2022 & ICASSP &160K & Audio & EER \\
 & LAV-DF~\cite{cai2022you} & 2022 & DICTA & 13K & Multimodal & ACC, AUC \\
 & Fake2M~\cite{lu2023seeing} & 2023 & NeurIPS & 20K & Image & ACC \\
 & DF-Platter~\cite{narayan2023df} & 2023 & CVPR & 133K & Video & FaceWA, FaceAuc, FLA, VLA\\
 & CFAD~\cite{ma2024cfad} & 2024 & SPEECH COMMUN & 374K & Audio & EER \\
 & MLAAD~\cite{muller2024mlaad} & 2024 & IJCNN & 76K & Audio & ACC \\
 & SVDD2024~\cite{zhang2024svdd} & 2024 & SLT & 84K & Audio & EER \\

\bottomrule
\end{NiceTabular}
}
\label{tab:deepfake_benchmarks}
\end{table*}

\subsubsection{Synthesis and governance implications}

Abuse and misuse defenses have evolved from single-modality detection toward multimodal analysis of text, image, audio, and video manipulation. However, this domain faces an asymmetric generalization problem: generative tools can rapidly produce novel synthetic content, while defensive systems must detect unseen models, modalities, and attack strategies. This creates a trade-off between detection coverage and false-positive risk, particularly in legal, journalistic, and public-safety contexts where incorrectly flagging legitimate content can also cause harm.

Deepfake and misuse governance therefore requires coordinated intervention across regulators, industry platforms, model providers, and downstream deployers. Although detection and watermarking are important technical countermeasures, effective governance also depends on provenance standards, disclosure requirements for AI-generated content, platform accountability policies, takedown and appeal procedures, and cross-border coordination. Because generative capabilities evolve quickly, governance systems should support continuous adaptation rather than rely only on static detection pipelines. Misuse-oriented red-teaming and abuse-case monitoring can help identify emerging patterns early and connect technical safeguards with enforceable content-integrity standards.
\hypertarget{Ethical security}{}
\section{Ethical security}\label{sec:social}
This section addresses the complex ethical dimensions emerging from the widespread deployment of AI technologies across critical societal domains. As AI systems increasingly influence human decision-making, healthcare, and education, they raise urgent questions about fairness, legal responsibility, and the moral status of both humans and non-human agents. We begin by examining how algorithmic decision-making can perpetuate social biases and affect public well-being. We then examine evolving ethical and legal frameworks. Finally, we discuss the distribution of responsibility across the AI lifecycle.

\hypertarget{Social and economic issues}{}
\subsection{Social and economic issues} \label{sec:economic}

\subsubsection{Problem definition}
The social and economic challenges of AI arise from its growing capacity to reshape labor markets, organizational decision-making, market competition, and public trust. At the social level, AI and automation may alter the demand for different types of labor, especially by reducing the need for routine or easily automated tasks while increasing demand for high-skilled and cognitive-intensive work. Such changes can intensify labor market polarization, job insecurity, and skill mismatch, particularly for workers in regions or sectors with limited access to retraining opportunities~\cite{xin2024robotics,fierro2022automation,lenzi2025income}. Older workers and other vulnerable groups may face additional risks when technological change outpaces institutional support for adaptation~\cite{alcover2021aging,gmyrek2023generative}. Beyond employment, AI systems can also affect social trust when algorithmic decisions reproduce historical bias or when personalized information systems reinforce selective exposure and echo chambers~\cite{lambrecht2019algorithmic,abrardi2022artificial}. Therefore, the social impact of AI is not limited to job displacement, but also includes fairness, inclusion, and the distribution of opportunities across different groups and communities.

At the economic level, AI can improve productivity and create new forms of value, but the magnitude and distribution of these gains remain context-dependent~\cite{wu2024artificial,corrado2021artificial}.
These productivity gains may be unevenly distributed across workers, firms, sectors, and regions. Geographic concentration is especially salient: prior Brookings analyses report that a small set of leading U.S. metro areas concentrated a large share of AI research, startup, and job-posting activity, and that six major metro areas accounted for 47\% of generative-AI job postings in a Brookings database~\cite{muro2023building}. AI may also reshape market competition. In the German retail gasoline market, algorithmic-pricing adoption increased margins by 9\% in non-monopoly markets, and margins increased by 28\% in duopoly markets where both competing stations adopted such software~\cite{assad2024algorithmic}. These findings suggest that AI-related economic security cannot be reduced to aggregate productivity gains alone, but requires governance mechanisms for inclusive labor transitions, regional development, fair competition, and continuous monitoring of distributional impacts.
\subsubsection{Existing methods}
Governance responses have emerged across technical, institutional, and regulatory domains to address the multifaceted risks of AI.

\noindent \fix{\textbf{(1) Technical measures.}}
Technical responses aim to mitigate AI’s societal and economic risks through innovations in algorithm design, privacy protections, and human-centered automation. These measures prioritize transparency, fairness, security, and adaptability.

\begin{itemize}

\item \fix{\textbf{Explainability and fairness mechanisms.}}
Technical levers include \emph{XAI for transparency}, \emph{privacy‑preserving learning}, \emph{human-centered robotics}, and \emph{adaptive reskilling platforms}. Together they target bias, privacy and displacement.

\item \fix{\textbf{Human-centered robotics and adaptive skills.}}
Robotics development increasingly emphasizes human-robot collaboration rather than substitution. Evidence from China suggests that robots deployed in rural areas can reduce labor market frictions and support employment among underrepresented groups \cite{xin2024robotics}. This aligns with calls for vocational training systems to adapt, equipping workers with skills that complement automation \cite{berg2023risks}.

\item \fix{\textbf{Responsive learning and long-term adaptation.}}
Automation often benefits high-skilled workers while displacing those in routine roles \cite{acemoglu2002technical}. Continuous learning systems and reskilling interfaces are essential to buffer such effects. Studies from Europe show that although robotics may reduce job intensity, employment quality can be preserved with adaptive interventions \cite{anton2023does}.

\end{itemize}

\noindent \fix{\textbf{(2) Policy and legal measures.}}
Policy frameworks and legal interventions address AI’s systemic risks by reforming market structures, enhancing social protections, and building institutional capacity for inclusive governance.

\begin{itemize}

\item \fix{\textbf{Labor protection and educational reform.}}
Redistributive mechanisms, such as capital taxation or universal basic income, are often proposed to mitigate displacement effects, although their effectiveness depends on design \cite{korinek2018artificial}. More targeted interventions include subsidies for training in “prediction-complementary” roles, which involve judgment tasks that AI cannot easily automate \cite{agrawal2018human}. Education systems must evolve to foster creativity, problem-solving, and empathy—skills that remain uniquely human \cite{autor2003skill}.

\item \fix{\textbf{Regional development and institutional responsiveness.}}
AI-driven growth tends to concentrate in major innovation hubs, leaving rural or peripheral regions behind \cite{choi2024artificial}. Regional innovation funds and fiscal incentives can support a more equitable distribution of AI benefits \cite{corradini2021geography}. At the same time, governments need adaptive labor market institutions to detect emerging skill mismatches through real-time analytics \cite{dahlke2024epidemic} and modernize social protections, such as wage insurance and flexible unemployment support.

\item \fix{\textbf{Implementation and evaluation capacity.}}
Effective AI governance hinges on timely implementation and rigorous evaluation. Policies must be forward-looking, sector-specific, and coordinated across labor, education, and industrial domains \cite{dosi2021embodied}. Iterative feedback mechanisms and evidence-based assessments are critical for refining strategies and ensuring long-term accountability \cite{rodrik2009industrial}.

\end{itemize}

\hypertarget{Ethical and legal issues}{}
\subsection{Ethical and legal issues} \label{sec:legal}

\subsubsection{Problem definition}

The ethical and legal challenges of AI arise from the increasing influence of AI systems on individual rights, social norms, and regulatory institutions. At the ethical level, AI systems may affect fairness, privacy, transparency, autonomy, and human dignity across high-impact domains such as healthcare, finance, education, employment, and public services. These concerns are closely connected to technical limitations: biased training data may produce unequal outcomes across demographic groups, opaque model behavior may hinder explanation and contestability, and insufficient human oversight may weaken accountability when automated decisions affect individuals or communities. Therefore, ethical security does not only concern whether an AI system is accurate or efficient, but also whether its development and deployment respect affected stakeholders, protect vulnerable groups, and remain open to scrutiny.

At the legal level, AI governance faces uncertainty because existing rules on liability, intellectual property, data protection, consent, and compliance differ across jurisdictions and continue to evolve with technological change. Risk-based regulatory frameworks, such as the EU AI Act, impose different obligations according to the risk level and use context of AI systems~\cite{eu2024aiact}. Copyright rules also remain contested, especially for AI-generated or AI-assisted content, where human authorship and creative control are central issues in current guidance~\cite{usco2023aiguidance}. Data protection regimes further differ in their treatment of consent, individual rights, and business obligations, as reflected by frameworks such as the GDPR and the CCPA~\cite{gdpr2016,ccpa2018}. This fragmentation makes it difficult to determine who should be responsible when AI systems cause harm, how generated content should be attributed, and what forms of documentation, auditing, or oversight are required. As a result, ethical and legal governance must combine technical safeguards with institutional mechanisms, including transparent documentation, accountability structures, human oversight, and adaptive regulatory compliance.

\subsubsection{Existing methods}
AI governance employs key methodologies, including Value-Sensitive Design (VSD), risk-based legal compliance, and technical tools that support auditing, documentation, and safeguards. These approaches form a layered strategy from value-sensitive design choices to regulatory obligations and operational evidence for accountability.

VSD integrates human values into AI system design. The FairPrism dataset shows its role in reducing bias in text generation \cite{fleisig2023fairprism}. In linguistics, VSD aligns AI with cultural norms \cite{bird2024must}. It also highlights developer well-being in high-stress fields, such as computer vision \cite{su2021affective}.

\fix{Legal compliance frameworks operationalize ethical principles through binding regulations.} Recent regulatory frameworks such as the EU AI Act further structure AI governance through risk tiers, including unacceptable-risk, high-risk, limited-risk, and minimal-risk systems \cite{eu2024aiact}. In particular, high-risk AI systems deployed in domains such as healthcare, education, employment, and critical infrastructure are subject to stricter requirements on robustness, transparency, human oversight, logging, and post-deployment monitoring \cite{eu2024aiact,nist2024genaiprofile}. These requirements increasingly connect governance objectives with technical mechanisms such as interpretability analysis, adversarial robustness evaluation, hallucination mitigation, and lifecycle auditing \cite{nist2024genaiprofile,iso202342001}.

Technical tools can support, but not replace, legal and ethical governance. For example, SynthASpoof contributes to privacy-preserving facial authentication \cite{fang2023synthaspoof}, EditGuard provides copyright verification for generative content \cite{zhang2024editguard}, FairCLIP addresses fairness concerns in vision-language models \cite{luo2024fairclip}, and LlavaGuard contributes to multimodal safety safeguards \cite{helff2024llavaguard}. \fix{Aetheria similarly uses multi-agent debate and collaboration to produce interpretable multimodal content-safety judgments and traceable audit reports~\cite{he2025aetheria}.} These tools are best understood as components of compliance and audit workflows rather than complete ethical-governance solutions.

Evaluation metrics also require careful interpretation. Metrics such as COMET \cite{rei2020comet} and BERTScore \cite{zhang2019bertscore} can complement BLEU by better capturing semantic adequacy in generated text, but they do not by themselves establish fairness; fairness-oriented evaluation still requires subgroup, language, and cultural-context analyses.

Several limitations remain. VSD can struggle with cross-cultural applicability \cite{bird2024must}, legal frameworks may lag behind technological change, and technical safeguards often translate imperfectly into legal duties. Overall, transparency, explainability, fairness assessment, and cross-domain deployment remain critical research gaps for ethical and legal AI governance.

\hypertarget{Responsibility and accountability mechanisms}{}
\subsection{Responsibility and accountability mechanisms}\label{sec:accountability}

\subsubsection{Problem definition}
\fix{As AI spreads into healthcare, finance, law enforcement, and autonomous driving, the question of \emph{who is responsible when things go wrong} becomes legally urgent and ethically fraught.} \fix{Modern systems increasingly span device--edge--cloud infrastructures and interacting models~\cite{an2026aiflow}, so diffuse control and black-box opacity often blur accountability.} Without clear governance, these risks erode public trust and slow beneficial adoption.

\noindent\textbf{(1) Role-based accountability across the AI lifecycle.}
For accountability mechanisms to function effectively, responsibilities must be defined across key roles \cite{deshpande2022responsible} in the AI lifecycle, including deployers, users, auditors, and regulators. \fix{\Cref{tab:roles} summarizes the duties and primary risks of six key stakeholders.}

\begin{table}[t]

\centering
\caption{Accountability across the AI lifecycle.}
\label{tab:roles}
\begin{tabular}{lll}
\toprule
\textbf{Role} & \textbf{Core duty} & \textbf{Primary risk}\\
\midrule
Designers  & Bias‑safe architecture; ethical‑by‑design          & Hidden bias\\
Developers & Secure code; full logs                             & Vulnerabilities\\
\fix{Deployers}  & \fix{Deployment approval; runtime monitoring}             & \fix{Operational misuse}\\
Users      & Due care; understand limits                        & Over‑reliance\\
Auditors   & Independent system checks                          & Capture and mis‑report\\
Regulators & Enforcement; redress                               & Regulatory lag\\
\bottomrule
\end{tabular}
\end{table}

\fix{Accountability allocation should be established before deployment rather than reconstructed only after an incident. A responsibility matrix should identify who owns model design, data stewardship, validation, deployment approval, runtime operation, independent audit, incident response, user redress, and regulatory notification. Each obligation should be linked to evidence---such as versioned test results, approval records, logs, and corrective-action reports---and to an escalation path when a threshold is exceeded. Because several actors may contribute to one outcome, shared responsibilities should be explicit without allowing collective involvement to become an accountability vacuum.}

\noindent \fix{\textbf{(2) Challenges in AI accountability.}}
While establishing clear roles and responsibilities is fundamental to AI governance, several systemic challenges complicate the practical implementation of accountability mechanisms.

\begin{itemize}

\item \fix{\textbf{Ambiguity in responsibility attribution.}}
AI system decisions are often the result of multi-party collaboration, including algorithm designers, developers, deploying institutions, and end users. This complex participation chain leads to a convoluted responsibility structure and increases the risk of unclear attribution. When AI decisions result in negative outcomes, involved parties may deflect blame, leading to an ``accountability vacuum" or "responsibility gap". As studies have pointed out \cite{wieringa2020account}, without a well-defined accountability framework, neither ethical responsibility nor legal liability can be assumed appropriately by relevant stakeholders.

\item \fix{\textbf{Responsibility shifting.}}
AI systems may be used to shift or dilute responsibility. On one hand, users may overly rely on AI decisions, transferring human responsibilities to machines. On the other hand, developers and deployers may use AI to evade their obligations. When AI systems make mistakes, people often blame ``the algorithm made a mistake'', reducing human responsibility \cite{diaz2023connecting}.

\item \fix{\textbf{Lack of transparency undermining accountability.}}
Many AI algorithms operate as ``black boxes'' — their decision-making processes are complex and lack interpretability. This lack of transparency makes it difficult to hold any party accountable \cite{falco2021governing}. When an accident or biased decision occurs, it is nearly impossible to determine what happened and why, especially without detailed logging and a traceable decision-making process.

\end{itemize}

\subsubsection{Existing methods}

Clear accountability frameworks combine \emph{technical safeguards} and \emph{policy levers} to ensure traceability, transparency, and enforceable liability.

\noindent\textbf{(1) Technical measures.}
\begin{itemize}
\item \textbf{Auditability and logging.}
To mitigate these risks, technical improvements aim to enhance the auditability and explainability of AI systems. This includes implementing comprehensive logging and audit-trail mechanisms that preserve key data and decision steps throughout a system’s operation. For instance, automated systems can incorporate audit tools like aviation ``black boxes'', recording high-fidelity data on system behavior and environmental context \cite{cen2024paths}. These audit trails provide critical post-event evidence, enabling independent analysts to reconstruct events, identify causes, and assign responsibility. \fix{Compliance tooling can automate evidence collection, control checks, and report generation, but the resulting output remains an input to an audit conducted under an institutionally defined scope, authority, independence requirement, and remediation process.}

\item \textbf{Traceability and explainability.}
Improving the traceability of AI decision-making ensures that the entire process — from input to model decision to output — can be tracked. This includes maintaining records of training data sources, model versions, and parameter changes. In the event of failure, these records enable the identification of the specific stage and party responsible.

\item \textbf{Continuous monitoring and incident reporting.}
Monitoring and alerting mechanisms should be deployed to capture AI anomalies or potential risks in real time. These systems provide crucial evidence before and after an incident. Such records offer valuable material for researchers and regulators to improve system design and reduce future risks. An open failure reporting mechanism encourages stakeholders to expose and fix problems promptly, rather than hide them to avoid responsibility.
\end{itemize}

\noindent\textbf{(2) Policy and legal measures.}

\begin{itemize}
\item \textbf{Accountability regulations and standards.}
Governments and industry groups are advancing legislation and standards for AI accountability. For example, the EU High-Level Expert Group published the ``Ethics Guidelines for Trustworthy AI'', identifying legality, ethical compliance, and technical robustness as core principles AI systems must meet, while emphasizing transparency and accountability. \fix{The EU AI Act establishes obligations for providers and deployers of high-risk AI systems \cite{eu2024aiact}.} Singapore’s AI governance framework also advocates for fairness, explainability, transparency, and human-centric practices across the AI lifecycle.

\item \textbf{Independent audits and certification.}
Independent third-party auditing systems are key to ensuring AI accountability. They help expose issues in decision-making and supervise stakeholders’ behavior. Scholars have proposed institutions like the Independent Auditing of AI Systems to audit highly automated systems and foster responsible development. Policymakers can require high-risk AI systems to pass qualification assessments or obtain licenses before deployment. Such external oversight pressures developers and deployers to follow safety and ethical norms. Audit institutions must also be held accountable. Industry associations or authorities should regulate their credentials, and misconduct such as falsified reports or collusion with audited entities should be punished. Proper oversight ensures independence and credibility in AI audits, preventing a regulatory vacuum. \fix{Automated compliance checks therefore support auditor judgment and traceability; they do not determine the normative acceptability of residual risk or substitute for governance bodies empowered to approve, restrict, or halt deployment.}

\item \textbf{Legal clarity and liability insurance.}
Legal frameworks must define the responsibilities of all stakeholders in the AI ecosystem to avoid blame-shifting. Without such clarity, disputes over responsibility are likely. Legal principles are needed to determine who is accountable for foreseeable and avoidable mistakes. Introducing liability insurance and compensation funds is another key strategy. Drawing from workplace injury compensation models, ``no-fault compensation'' systems can enable victims of AI-related harm to be compensated swiftly — without lengthy fault-finding procedures. This guarantees redress for victims and encourages developers and users to report problems and learn from them without fear of litigation. When combined with mandatory incident reporting and independent investigative institutions, a closed-loop system of accountability and continuous improvement can be formed.
\end{itemize}
\hypertarget{Open challenges and future directions}{}
\section{Discussion and outlook} \label{sec:future}

The preceding sections have reviewed AI governance through a three-layer taxonomy covering intrinsic model reliability, derivative risks arising in deployment contexts, and ethical or institutional mechanisms for accountability. This section synthesizes the cross-cutting implications of this taxonomy. Specifically, it identifies recurring technical limitations across the surveyed domains, discusses the role and limitations of benchmark-based evaluation in governance workflows, and outlines research directions for developing AI systems whose safety, privacy, fairness, and accountability can be assessed throughout the lifecycle.

\hypertarget{Technical gaps and structural trade-offs}{}
\subsection{Technical gaps and structural trade-offs} \label{sec:gaps}

Across the domains reviewed in this survey, the most persistent technical gaps are not isolated failures of individual methods. They repeatedly reflect limitations in generalization, deployment transfer, and the balancing of competing governance objectives. We summarize these gaps below according to the main risk categories discussed in Sections~\ref{sec:intrinstic}--\ref{sec:social}.

\begin{itemize}

\item \fix{\textbf{Insufficient adversarial robustness.}}
Adversarial robustness remains a critical barrier to secure AI deployment, as current defenses fail to generalize against evolving attack vectors, particularly in multimodal systems. \fix{Cross-modal manipulation can generate adversarial examples by perturbing visual--textual alignment or by combining individually benign inputs into unsafe multimodal reasoning chains~\cite{cai2026safeunimodal}.} \fix{Existing methods, such as prompt-based robust tuning \cite{li2024one}, remain constrained by specific threat models and can struggle against unrestricted attacks, such as adversarial patches \cite{brown2017adversarial}.} A central challenge is therefore not only to design stronger defenses, but also to evaluate whether defenses remain effective under adaptive, cross-modal, and deployment-specific threat models.

\item \fix{\textbf{Persistent hallucinations in LLMs and MLLMs.}}
Hallucinations, defined as plausible but incorrect outputs, reduce reliability of LLMs and MLLMs in critical fields like healthcare and education. These issues stem from model flaws and biased data, with current methods like post-hoc validation (\eg, Silkie, VIGC, Woodspecker \cite{wang2024vigc}) having limited effect (\cref{sec:hallu}). Future work should move beyond isolated factuality improvements toward reliability mechanisms that combine uncertainty estimation, retrieval or tool-based verification, domain-specific knowledge constraints, and user-feedback loops. This is especially important for high-risk scenarios where the cost of a fluent but incorrect answer is substantially higher than the cost of abstention or escalation.

\item \fix{\textbf{Limited interpretability of black-box models.}}
Limited interpretability reduces transparency and accountability of black-box models, especially in sensitive areas like healthcare \cite{luo2024croup}. The survey notes attribution methods, such as Grad-CAM \cite{selvaraju2017grad} and Integrated Gradients \cite{sundararajan2017axiomatic}, offer partial insights but cannot fully explain complex decisions, particularly in multimodal settings. A key open challenge is to make interpretability useful for governance rather than merely explanatory visualization. Interpretable evidence should support auditability, failure diagnosis, contestability, and responsibility assignment for different stakeholders, including developers, domain experts, regulators, and affected users.

\item \fix{\textbf{Privacy vulnerabilities in data-intensive models.}}
Privacy risks remain in LLMs (\cref{sec:privacy}). Differential privacy and federated learning help but have trade-offs, as noise hurts performance and federated systems are still open to attacks like membership inference \cite{li2025generating} and model inversion \cite{carlini2021extracting}. Personalized applications, such as LLM agents and retrieval-augmented generation systems, further expose sensitive data \cite{zhong2025rtbas}. Secure multi-party computation \cite{hou2023ciphergpt} with lightweight models may enable scalable privacy-preserving inference, but privacy protection must be evaluated together with utility, latency, deployment feasibility, and data-governance requirements such as consent, retention, and access control.

\item \fix{\textbf{Bias propagation in model outputs.}}
Bias propagation in AI systems, particularly in computer vision and LLMs, perpetuates unfair outcomes across demographic groups, as discussed in~\cref{sec:bias}. Benchmarks like VLBiasBench and FACET reveal disparities in performance related to gender and race, often stemming from biased training data \cite{wang2024vlbiasbench}. Fairness-aware training procedures, such as demographic parity constraints \cite{zafar2017fairness}, can mitigate biases, but dataset imbalance and context-specific fairness definitions limit their effectiveness. Future progress requires fairness evaluation that is sensitive to intersectional groups, deployment context, language and cultural variation, and the trade-off between group-level parity and task-specific utility.

\item \fix{\textbf{Inadequate generalization in misuse detection.}}
Misuse prevention remains difficult because harmful uses of AI systems evolve together with model capabilities. Deepfake detectors, such as CNN-based models (\eg, AASIST \cite{jung2022aasist}) and multimodal fusion methods (\eg, AVFakeNet \cite{ILYAS2023110124}), often struggle against unseen synthetic-media generation techniques based on advanced GANs or diffusion models. Similar generalization problems arise in broader AI-enabled misuse, including jailbreaks, tool-mediated abuse, and automated misinformation. Robust detection therefore requires cross-modal evidence, adaptive evaluation, and deployment-time monitoring rather than reliance on a fixed set of known attack patterns.

\end{itemize}

These domain-specific gaps reveal a broader structural issue: AI governance must manage trade-offs among safety, utility, privacy, fairness, computational cost, and accountability. Some tensions are resource-based. \fix{For example, prompt-based robust tuning can improve VLM robustness, but such defenses remain computationally expensive or constrained by specific threat models \cite{li2024one}.} Similarly, differential privacy and privacy-preserving prompt tuning reduce leakage risks, but must manage utility loss during fine-tuning \cite{behnia2022ew,li2023privacy,du2023dp}; secure inference systems such as CipherGPT and PUMA protect user inputs and model parameters, but introduce additional inference and deployment overhead \cite{hou2023ciphergpt,dong2023puma}. \fix{Generative transmission systems expose a related system-level trade-off by jointly balancing bandwidth, computation, memory reuse, weak-network robustness, and perceptual quality~\cite{chen2026gentrans}.}

Other tensions are normative as well as technical. Uncertainty-aware hallucination mitigation and abstention mechanisms can improve factual reliability, but they require decisions about when a system should answer, refuse, or defer \cite{farquhar2024detecting,zhang2024vl}. Fairness-constrained optimization can reduce group-level disparities, but it requires domain-specific choices about which disparities matter and how performance differences should be judged \cite{zafar2017fairness}. These examples show why governance cannot be reduced to maximizing a single benchmark score. Instead, acceptable operating points must be specified for each deployment domain through documented safety, utility, privacy, and fairness requirements, followed by post-deployment audits that monitor whether these trade-offs drift as data, users, and threat models change.

\hypertarget{Governance-aware evaluation}{}
\subsection{Governance-aware evaluation} \label{sec:dynamic-governance-evaluation}

The benchmark tables in this survey provide structured resources for evaluating adversarial robustness, hallucination, privacy, fairness, and misuse. A natural question is how these benchmark results should be used once AI systems move from controlled evaluation settings to real deployment environments. Static academic benchmarks such as AdvGLUE~\cite{wang2021adversarial_glue}, TruthfulQA~\cite{lin2022truthfulqa}, privacy benchmarks, fairness datasets, and adversarial safety suites are valuable because they enable reproducible comparison. However, they are usually constructed offline with fixed samples, predefined labels, and frozen evaluation settings, whereas deployed AI systems operate under changing user behavior, deployment contexts, social norms, data distributions, and adversarial strategies.

For this reason, benchmark scores should be interpreted as part of a governance evidence package rather than as final compliance proof. Industrial entry standards may use benchmark thresholds as minimum requirements, but they should also specify acceptable degradation under distribution shift, worst-case subgroup performance, refusal behavior, privacy leakage tolerance, and incident-response obligations. For high-risk systems, a single pass on a benchmark should not be treated as a permanent license to deploy; evaluation should be repeated after model updates, data refreshes, prompt-template changes, fine-tuning, or shifts in the user population \cite{eu2024aiact,nist2024genaiprofile}.

Governance-aware evaluation should also move beyond separate risk-specific tests toward joint evaluation under realistic deployment scenarios. In practice, risks are often coupled rather than independent: privacy-preserving training may affect subgroup performance, hallucination control may change refusal behavior and user utility, and robustness interventions may alter calibration or downstream task performance. A minimal joint-evaluation workflow can include four steps: scenario-based testing to define high-stakes deployment cases; multi-metric profiling to report robustness, factuality, privacy leakage, fairness disparity, calibration, and misuse resistance side by side; trade-off analysis to identify where improving one dimension degrades another; and governance review to translate the resulting profile into deployment thresholds, residual-risk documentation, monitoring requirements, and responsibility assignment.

This evaluation process also provides a practical way to operationalize the three-layer framework proposed in this survey. \fix{All three security layers should be assessed at each lifecycle stage, although their principal artifacts and responsible actors may differ.} \fix{During development, intrinsic-security requirements should populate the threat model, risk register, evaluation plan, and model-card evidence for adversarial vulnerability, distribution shift, hallucination, and interpretability.} \fix{Before release, derivative-security findings should be recorded in an audit report and deployment checklist covering privacy leakage, subgroup performance, misuse testing, residual risks, and named sign-off owners; the resulting model or system card should state intended uses, limitations, thresholds, and escalation conditions.} \fix{During operation, ethical-security controls should be implemented through monitoring dashboards, audit logs, incident reports, change-control records, and periodic review by accountable institutional roles.} \fix{For example, an LLM-based healthcare triage assistant should not be approved solely on aggregate accuracy; its deployment checklist should also require uncertainty handling, privacy-preserving processing of patient information, subgroup-performance thresholds, responsibility allocation, escalation routes, and post-deployment monitoring.}

\fix{Accordingly, these artifacts should form an auditable evidence chain rather than a collection of one-off documents: benchmark and red-team results feed the audit report; the audit report informs the deployment checklist and system card; approved thresholds configure post-deployment monitoring; and incidents or material model changes trigger re-evaluation and updated documentation.} \fix{Dynamic red-teaming can continuously probe new jailbreaks, misinformation tactics, privacy extraction strategies, and multimodal manipulation patterns \cite{debenedetti2024agentdojo,ruan2024toolemu,nist2024genaiprofile}.}

\hypertarget{Regulatory and ethical considerations}{}
\subsection{Regulatory and ethical considerations} \label{sec:consideration}

Beyond technical evaluation, AI governance also depends on regulatory and ethical coordination across jurisdictions, cultures, and institutional roles. Three recurring concerns are especially important for translating technical safeguards into enforceable and socially legitimate governance mechanisms.

\begin{itemize}

\item \fix{\textbf{Regulatory fragmentation across jurisdictions.}}
Regulatory fragmentation hinders global AI governance, as differing legal systems, cultural norms, industrial priorities, and public-risk perceptions create inconsistent compliance requirements. The survey highlights frameworks such as \cite{Sarridis2025FLAC}, which promote transparency but vary in their applicability. Rather than attempting to impose a single universal rule, a more practical direction is to develop modular regulatory frameworks that preserve shared principles, such as risk-based classification, transparency, human oversight, auditability, and post-deployment monitoring, while allowing context-specific adaptation across jurisdictions.

\item \fix{\textbf{Cultural gaps in ethical guidelines.}}
Cultural gaps in ethical guidelines limit their applicability, particularly in addressing implicit and intersectional biases \cite{bai2025explicitly}. The survey notes that benchmarks like VLBiasBench and FACET are often Western-centric, failing to capture global diversity \cite{wang2024vlbiasbench,gustafson2023facet}. Developing multilingual and multicultural ethical guidelines requires benchmarks and evaluation protocols that reflect diverse social contexts, languages, and institutional expectations. Such guidelines should also remain revisable through public feedback, case-based learning, and evidence from real deployment failures.

\item \fix{\textbf{Accountability gaps in AI deployment.}}
Accountability gaps complicate enforcement, as the survey notes the lack of standardized mechanisms for assigning responsibility. \cref{sec:accountability} highlights the need for precise responsibility mapping across designers, developers, deployers, users, auditors, and regulators. Standardized auditability and logging mechanisms can preserve decision trails and enable post-hoc analysis. Interdisciplinary frameworks that integrate legal and technical accountability measures, such as third-party audits, incident reporting, and liability allocation, would strengthen enforcement and maintain societal legitimacy.

\end{itemize}

\hypertarget{Future research directions}{}
\subsection{Future research directions} \label{sec:opportunities}

The discussion above suggests several research directions that are closely tied to the main structure of this survey: improving the generalization of technical safeguards, embedding governance requirements into model development, building evaluation resources that reflect global deployment contexts, and strengthening collaboration between technical and institutional actors.

\begin{itemize}

\item \fix{\textbf{Generalizable defense mechanisms.}}
Future defenses should be evaluated not only by their performance against known attacks, but also by their transferability to adaptive, unseen, and multimodal threats. This requires defense mechanisms that combine training-time robustness, inference-time monitoring, uncertainty estimation, and incident-driven updating. For deployment, lightweight and real-time defenses are particularly important in settings such as online content moderation, AI agents, and safety-critical perception systems.

\item \fix{\textbf{Governance by design in AI development.}}
Robustness, privacy, fairness, and accountability should be incorporated into the design and training stages rather than treated as post-hoc corrections \cite{hendrycks2019benchmarkingCorruptions}. Multi-objective optimization can help balance robustness, performance, privacy, and efficiency, but technical optimization alone is insufficient. Future systems should also document design assumptions, supported deployment domains, known limitations, and re-evaluation triggers, so that governance requirements are embedded into the development lifecycle.

\item \fix{\textbf{Global and multicultural evaluation.}}
Comprehensive benchmarking remains crucial for evaluating AI governance, as current benchmarks, such as VLBiasBench, FACET, and TruthfulQA, still lack sufficient coverage across modalities, languages, cultures, and deployment contexts \cite{wang2024vlbiasbench}. Future benchmarks should include global, multilingual, and multicultural scenarios, while also supporting dynamic updates and user-feedback mechanisms. These benchmarks should complement, rather than replace, the governance-aware evaluation process discussed above.

\item \fix{\textbf{Cross-disciplinary governance infrastructure.}}
Cross-disciplinary collaboration is essential for addressing systemic governance challenges in domains such as healthcare, law, education, and public services. Domain expertise can help identify realistic failure modes, define acceptable trade-offs, and interpret evaluation results in context. Future work should develop co-design frameworks, shared audit protocols, and education programs that connect technical researchers, policymakers, legal experts, domain practitioners, and affected communities.

\end{itemize}

\section{Conclusion}
This paper offers a comprehensive overview of AI governance, addressing challenges across intrinsic security, derivative security, and ethical security. As AI systems permeate critical sectors like healthcare, education, and public policy, their risks, which range from adversarial attacks and privacy breaches to bias and societal impacts, demand governance frameworks that ensure transparency, accountability, and fairness. Our survey advocates for an integrated approach that balances technical robustness with ethical responsibility, emphasizing interdisciplinary collaboration to refine evaluation metrics, strengthen global standards, and guide responsible AI deployment. Continued research and policy development are essential to building AI systems that are secure, equitable, and aligned with public interests.

\Acknowledgements{This work was supported by the National Natural Science Foundation of China (Grant Nos. 62476224, 62472046, 62576046, 62406028, 62406021, 62336001, T2541022); the Young Elite Scientists Sponsorship Program of the Beijing High Innovation Plan (Grant No. 20250866); the Beijing Academy of Artificial Intelligence (Grant No. Z251100008125041); and the Frontier Technologies R\&D Program of Jiangsu (Grant No. BF2025012). The authors thank the anonymous reviewers for their valuable comments and suggestions to improve the quality of this paper.}

\bibliographystyle{scis}
\bibliography{references}

\end{document}